\documentclass{iopart}
\usepackage{graphicx}
\usepackage{latexsym}
\usepackage{iopams, amsopn, amstext}
\usepackage{colonequals}
\usepackage{dcolumn}
\usepackage{setspace}
\usepackage{subfigure}
\usepackage{bm}
\usepackage{url}
\usepackage{citesort}
\usepackage{rotating}
\usepackage{braket}
\usepackage{units} %  do things like \units[1.234]{sec}
\usepackage[usenames,dvipsnames]{color}

\usepackage{ulem}
\begin{document}

\title[Numerical simulations of compact object binaries]{Numerical
  simulations of compact object binaries}

\newcommand{\CITA}{\address{Canadian Institute for Theoretical
    Astrophysics, University of Toronto, 60~St.~George Street,
    Toronto, Ontario M5S 3H8, Canada}} %

\author{Harald P.\@ Pfeiffer}\CITA

\begin{abstract}
  Coalescing compact object binaries consisting of black holes and/or
  Neutron stars are a prime target for ground-based gravitational wave
  detectors. This article reviews the status of numerical simulations
  of these systems, with an emphasis on recent progress.
  %  This article reviews the current state and recent new
  %results in fully relativistic numerical simulations of compact
  %object binaries consisting of Black Holes and or Neutron Stars.  
\end{abstract}

\maketitle

\section{Introduction}
\label{sec:introduction}

Inspiraling and coalescing compact object binaries consisting of black
holes and/or Neutron stars are of high importance to understand
gravity and matter in extreme conditions.  The strong gravitational
fields require full nonlinear general relativity to describe them, and
so compact object binaries allow exploration of curved space-time in the 
nonlinear and highly dynamic regime.  The matter density inside
Neutron stars is above nuclear density.  Neutron stars therefore
provide an avenue to study matter at super-nuclear densities, and
compact object mergers allow such studies even in dynamic
situations when the two objects collide.

Compact object binaries are therefore at the centre of
attention for several scientific disciplines: The study of black hole
binaries (BH-BH) sheds light on properties of general relativity in
the genuinely nonlinear, dynamic regime (in a clean vacuum
environment), ranging from black hole kicks
(e.g.~\cite{Lousto:2011kp}) to the topological structure of event
horizons~\cite{Cohen:2011cf,Cohen:2008wa,Ponce:2010fq}.  Simulations
of binaries involving one or two Neutron stars (BH-NS, NS-NS)~\cite{Duez:2009yz,lrr-2011-6}
elucidate the connection between these systems and 
short gamma ray bursts.  
Finally, gravitational waves emitted by compact object binaries are the most
promising source for gravitational wave detectors like Advanced LIGO~\cite{Abbott:2007kv,Shoemaker:aLIGO,2010CQGra..27h4006H},
Advanced Virgo~\cite{Accadia:2011zz,aVIRGO} and KAGRA~\cite{Somiya:2011me}.
%The most urgent and arguably most
%compelling motivation to study
%compact object binaries arises from gravitational wave detectors.
Gravitational wave detectors require accurate waveform models
and information about electromagnetic
counterparts to optimally search for gravitational waves via matched
filtering (``event detection''), and to extract the maximum amount of
information about the source of gravitational waves, once such waves
have been detected (``parameter estimation'').

Direct numerical simulations of the full Einstein equations are an important
cornerstone in the study of coalescing compact object binaries. 
These simulations have continued to progress swiftly, resulting in a
large number of notable recent results.  This
article gives an overview of the current status of numerical
simulations of compact object binaries.  Our particular focus lies on
recent advances, which have not yet been incorporated into longer or more specialized
review articles (like~\cite{Centrella:2010mx,lrr-2011-6,McWilliams:2010iq}) 
or books
(like~\cite{Baumgarte-Shapiro-Book:2010}).  Numerical simulations of
higher-dimensional gravity (e.g.~\cite{Witek:2010xi,Lehner:2010pn}) and
alternative theories of gravity
(e.g.~\cite{Healy:2011ef,Paschalidis:2011ww}) are advancing rapidly,
but because of length limitations we shall restrict our attention to
the ``classical'' compact object binaries (BH-BH, BH-NS, NS-NS) in
four space-time dimensions in standard general relativity, the
scenario of most direct relevance to gravitational wave detectors.

This article is organized as follows:  Section~\ref{sec:bhbh} reviews 
BH-BH simulations, starting with numerical methods, recent advances,
applications to gravitational wave science, and ending with BH-BH
simulations embedded in gaseous or electromagnetic environments.
Section~\ref{sec:BH-NS/NS-NS} deals with binaries with Neutron stars
(BH-NS, NS-NS).  We conclude in Sec.~\ref{sec:discussion} with some
thoughts about the future of the field.

\section{Black Hole---Black Hole Binaries}
\label{sec:bhbh}

\subsection{Numerical methods}
\label{sec:NumericalMethods}

Since the numerical relativity breakthroughs in
2005~\cite{Pretorius:2005gq,Campanelli:2005dd,Baker:2005vv}, several
frameworks have emerged to simulate inspiraling and merging BH--BH
binaries.  The major differentiating factors are the choices for the
formulation of Einstein's equations (either the BSSNOK
system~\cite{Nakamura:1987zz,Shibata:1995we,Baumgarte:1998te} or the
Generalized Harmonic (GH)
system~\cite{Friedrich1985,Pretorius:2004jg,Pretorius:2005gq,Lindblom:2005qh})
and the choice of the numerical evolution algorithm,
finite-differences (FD) or pseudo-spectral methods.  Pretorius' first
simulations employed the combination
GH+FD~\cite{Pretorius:2004jg,Pretorius:2005gq}, and this code has since 
been applied to a variety of physical scenarios including high energy
collisions of BH--BH~\cite{Sperhake:2008ga}, eccentric
BH--BH~\cite{Pretorius:2007jn} and studies of gravity in different
space-time dimensions~\cite{Lehner:2010pn,Ramazanoglu:2010aj}.

%%%%%%%%%%%%%%%%%%%%%%%%%%%%%%%%%%%%%%%%%%%%%%%%%%%%%%%%%%%%%%%%
\begin{table}
\begin{tabular}{|p{2.64cm}||p{4.5cm}|p{4.5cm}|}
\hline 
&  BSSNOK+FD & GH+Spectral \\ \hline \hline
\multicolumn{3}{|l|}{\hspace*{2cm} \bf \emph{Initial data}}\\ \hline
Formulation & Conformal method using $\;\;\;\;$ Bowen-York solutions~\cite{O'Murchadha:1974nc,Pfeiffer:2002iy,Bowen:1980yu} & Conformal thin sandwich $\;\;$ \cite{York:1998hy,Pfeiffer:2002iy}  \\ \hline
Treatment of BH singularities & puncture data~\cite{Brandt:1997tf} &  quasi-equilibrium $\quad\qquad$ BH excision~\cite{Caudill:2006hw,Cook:2004kt,Cook:2001wi} \\ \hline
Num. algorithm & pseudo-spectral~\cite{Ansorg:2004ds} & pseudo-spectral~\cite{Pfeiffer:2002wt} \\ \hline
Low-eccentricity orbital param's & post-Newtonian inspiral \cite{Husa:2007rh}%, iterative ecc. removal~\cite{Purrer:2012wy,Tichy:2010qa}
 & iterative eccentricity removal \cite{Pfeiffer:2007yz,Buonanno:2010yk} \\ \hline \hline
\multicolumn{3}{|l|}{\hspace*{2cm}\bf \emph{Evolution}} \\ \hline
Formulation & BSSNOK~\cite{Nakamura:1987zz,Shibata:1995we,Baumgarte:1998te}
 & Generalized harmonic \& constraint damping~\cite{Friedrich1985,Pretorius:2004jg,Pretorius:2005gq,Lindblom:2005qh} \\ \hline
Gauge conditions & evolved lapse and shift $\quad\;\;\;$ \cite{Bona:1997hp,Alcubierre:2002kk,vanMeter:2006vi} &  Harmonic ($H^\mu\!=\!0$) $\qquad$ and/or 
 evolved $H^\mu$~\cite{Szilagyi:2009qz} \\ \hline
Treatment of BH singularities & moving punctures~\cite{Campanelli:2005dd,Baker:2005vv} & BH excision \cite{Scheel:2006gg} \\ \hline
Treatment of outer boundary & Sommerfeld BC & minimally--reflective, constraint--preserving \cite{Lindblom:2005qh,Rinne:2007ui} \\ \hline
Discretization & high-order finite-differences$\;\;\;$ \cite{Husa:2007hp,Zlochower:2005bj} & 
pseudo-spectral methods \\   \hline
Mesh-refinement & adaptive mesh refinement & domain decomposition\cite{Pfeiffer:2002wt,Szilagyi:2009qz} 
\\ \hline \hline
\multicolumn{3}{|l|}{\hspace*{2cm}\bf \emph{Diagnostics}} \\ \hline
BH & \multicolumn{2}{c|}{Apparent horizon finder, quasi-local spin measures } \\ \hline
GW extraction & \multicolumn{2}{c|}{Newman Penrose scalars, Regge-Wheeler-Zerilli} 
\\ \hline \hline
\multicolumn{3}{|l|}{\hspace*{2cm}\bf \emph{Major codes}} \\ \hline
Infrastructure & BAM, Cactus, Einstein $\qquad$ Toolkit, Hahndol & SpEC \\ \hline
Codes & BAM, Hahndol, LazEv, Lean, Llama, MayaKranc, UIUC  & SpEC \\ \hline
\end{tabular}
\caption{\label{tab:1}Ingredients into a BH-BH simulation, and the
most common choices for the two major frameworks in use to solve BH-BH spacetimes.}
\end{table}
%%%%%%%%%%%%%%%%%%%%%%%%%%%%%%%%%%%%%%%%%%%%%%%%%%%%%%%%%%%%%%%%

However, in regard to the main focus of this section, BH--BH simulations of
direct relevance to gravitational wave detectors, the combinations BSSNOK+FD and
GH+Spectral have emerged as the leading frameworks, and so we shall
describe these in some detail.  The BSSNOK+FD framework is built
around the
BSSNOK~\cite{Nakamura:1987zz,Shibata:1995we,Baumgarte:1998te}
evolution equations coupled with finite-difference (FD) approximation
schemes.  The GH+Spectral framework utilizes the generalized harmonic
(GH)
equations~\cite{Friedrich1985,Pretorius:2004jg,Pretorius:2005gq,Lindblom:2005qh}
and multi-domain spectral methods.  These frameworks differ in a large number of
the individual elements that are required to compute a gravitational
waveform with numerical relativity.  Table~\ref{tab:1} lists the main
choices that are made within the two frameworks for setting initial
data, performing the evolution and analyzing the resulting data.
Besides formulation and numerical algorithm used for the evolution equations,
these choices include formulation and numerical methods for the initial data, diagnostics tools, and 
%These choices include the formulation used to express Einstein's
%equations as well-posed partial differential equations~\cite{Gundlach:2006tw,Ce%ntrella:2010mx}, the numerical
%algorithms employed to solve the partial differential equations, and
factors that determine the physics of the scenario under consideration
(e.g. low orbital eccentricity).  
While Table~\ref{tab:1} conveys the
broad picture, it must necessarily simplify and omit details.  For
instance, Pretorius' breakthrough work~\cite{Pretorius:2005gq} does
not fit into these categories\footnote{Pretorius used the generalized
  harmonic evolution system with an adaptive-mesh-refinement
  finite-difference code, starting from initial data that contained
  nonsingular balls of scalar field that collapsed to BH's early in
  the evolution.}.

The choices shown in Table~\ref{tab:1} are interrelated
and only certain combinations are feasible.   In particular, one 
cannot simply exchange specific elements between BSSNOK+FD and GH+Spectral.
Let us illustrate some of these dependencies: 
Moving puncture evolutions performed in the BSSNOK+FD
framework require very specific gauge conditions (Gamma-driver and
1+log slicing~\cite{Bona:1997hp,Alcubierre:2002kk,vanMeter:2006vi}) to
keep the punctures stable during the evolution. 
Because the BH's are not excised,
moving puncture evolutions require initial data covering the entire
initial--data hypersurface.  Therefore, a moving puncture evolution
cannot be started from the excision initial--data used for the spectral
evolutions\footnote{Proposals to fill-in the BH
  interiors of excision initial--data~\cite{Brown:2007pg,Brown:2008sb,Etienne:2007hr} have
  not been systematically pursued.}.

Puncture data covers the entire initial-data hypersurface, so it seems that GH+Spectral evolutions could use puncture data.  However, this is also non-trivial: 
Singularity excision requires careful control over the
location and shape of the excision
boundary~\cite{Scheel:2006gg,Lovelace:2011nu}.  This is more difficult when the black hole horizons change on short time-scales, in particular during merger\footnote{Control of the excision boundary was one of the
major obstacles in obtaining BH--BH mergers with GH+Spectral, cf.~\cite{Scheel:2006gg,Lovelace:2011nu}.} and early in an evolution when the coordinate location 
of the black holes and the gauge settles down.
Conformal thin
sandwich initial data (as used in the GH+Spectral approach) 
reduces significantly such transients and is therefore more suitable
for GH+Spectral than puncture initial data.

Trumpet initial data reduces initial transient apparent horizon motion~\cite{Hannam:2009ib}
compared to standard puncture initial data, however, so far trumpet
initial data not been used regularly for production BH--BH simulations.

To summarize some key properties of the two BH--BH frameworks:
\begin{itemize}
\item BSSNOK+FD simulations are generally considered more robust and
  ``easier'' because of the lack of black hole excision and the larger
  expertise of how to stabilize finite-difference methods.  Moreover,
  code-infrastructure~\cite{Goodale02a,cactus,Loffler:2011ay},
 including
  adaptive-mesh-refinement (Carpet~\cite{Schnetter:2003rb,carpet}) and
  apparent horizon finders~\cite{Thornburg:2003sf},
  are publicly available.  Several independent research groups use
  these tools and have obtained a tremendous wealth of results
  (e.g. black hole kicks).  The major codes are listed in
  Table~\ref{tab:1} and more details are available in the related
  proceeding contributions describing the NINJA collaboration~\cite{Ajith:2012tt}.
\item Spectral methods, in contrast, require BH excision and are very
  sensitive to the presence of ill-defined elements in the problem (e.g. outer
  boundary conditions, handling of inter-domain boundaries where grids
  of different resolution touch, or filtering).  Significant care and
  expertise is needed to analyze and control all relevant aspects of
  the evolution, and only one spectral BH-BH code exists: The Spectral
  Einstein Code {\tt SpEC}~\cite{SpECwebsite}, developed by the SXS
  collaboration between Cornell, Caltech, CITA and Washington State University.
\item The key advantage of spectral methods is their fast (exponential)
  convergence.  Therefore, high accuracy can be obtained with
  comparatively low computational cost.  The high accuracy/lower cost
  permeates the entire ``simulation pipeline'': Lower numerical noise
  (due to higher accuracy) makes it easier to perform certain fits 
that are required in iterative eccentricity removal (see Sec.~\ref{sec:techniques}); the low computational cost makes it possible to perform
  longer simulations; the clean waveforms extracted at finite radius
  are more amenable to extrapolation to obtain the asymptotic waveform
  at infinitely large distance; high accuracy also makes it easier  to compute and
  analyze quantities that depend on higher derivatives of the evolved
  fields, for instance black hole vortices and
  tendices~\cite{Owen:2010fa,Nichols:2011pu}.
\item To date, many more simulations have been performed
  with BSSNOK+FD than with GH+Spectral, but the GH+Spectral
  simulations are longer and more accurate.  Both
  techniques have found their place, depending on the precise needs of
  the science to be studied.  However, this conventional wisdom slowly
  changes, as the BSSN+FD simulations become more accurate and achieve
  longer simulations, whereas the GH+Spectral simulations become more
  robust and the number of GH+Spectral simulations 
  increases~\cite{Ajith:2012tt}.
\end{itemize}

While simulations work successfully, we caution the reader that
both frameworks utilize heuristically determined
ingredients: Gauge conditions and constraint damping terms contain
parameters chosen by trial and error, and mesh structures are tuned based
on user experience.  Therefore, parameter choices working in one region of
parameter space may require adjustments for other regions
(e.g.~\cite{Muller:2010zze,Schnetter:2010cz,Alic:2010wu}).  Furthermore, 
constraint
damping parameters impact the accuracy and quality of the
simulation beyond their main purpose of preserving the
constraints~\cite{Chu:2009md}.  The presence of user-tuned parameters
has further implications: On the negative side, it is not guaranteed
that the current techniques work in unexplored regions of parameter
space.  On the positive side, it is conceivable --even likely-- that further tuning of code
parameters will enhance accuracy and efficiency of future simulations compared to todays runs.

%%%%%%%%%%%%%%%%%%%%%%%%%%%%%%%%%%%%%%%%%%%%%%%%%%%%%%%%%%%%%%%%

\subsection{Recent advances}
\label{sec:recent-advances}

Given the rather mature state of BH-BH simulations, recent advances
are incremental, pushing and refining capabilities, and confirming
assumptions that were made in previous years to achieve fast progress.

\subsubsection{New records}

Let us start with some ``new records'': With increasing mass-ratio
$q\!=\!M_1/M_2\ge 1$, fully numerical simulations become more challenging
because the computational cost increases (the small black hole must be
resolved), and because experience of code tuning from comparable mass
binaries is no longer applicable.  Two groups  pushed the
mass-ratio of BH-BH simulations up to $q\!=\!100$: Sperhake et
al~\cite{Sperhake:2011ik} examine head-on collisions of a small black
hole with a large black hole.  Lousto and
Zlochower~\cite{Lousto:2010ut} perform a numerical simulation of the
last two orbits of a BH-BH binary of mass-ratio $q\!=\!100$.  Both these
calculations are very impressive, requiring not only a large amount of
CPU resources, but also improvements and tuning of the computational
infrastructure (e.g. location of mesh-refinement regions).  The RIT
group~\cite{Nakano:2011pb,Lousto:2010qx,Lousto:2010tb} extracts
BH--trajectories from numerical simulations of mass-ratio up to
$q\!=\!100$.  These trajectories are then used in perturbative
calculations to compute the waveforms of high mass-ratio binaries.

Turning to spin, the long-standing bound 
\begin{equation}\label{eq:spin0.93}
\frac{S}{M^2}\equiv\chi\lesssim 0.93
\end{equation} 
was recently broken.   Certain black hole properties vary significantly between $\chi\!=\!0.93$ and $\chi\!=\!1$ and therefore angular momentum may not be the most useful measure of extremality.  For instance, the rotational energy of a black hole with $\chi\!=\!0.93$ is only $59\%$ of the rotational energy of a maximally rotating black hole.  When taking rotational energy as the metric, the bound~(\ref{eq:spin0.93}) is far from extremality.

Equation~(\ref{eq:spin0.93}) represents the maximal spin that is achievable with puncture initial data~\cite{Dain:2008ck}, and
therefore, all BH-BH simulations within the BSSN-OK framework are
limited by this bound.  BH spins higher than the
bound~(\ref{eq:spin0.93}) are achieved with the more general
initial-data formalism utilized within the GH+Spectral framework.
Lovelace et al~\cite{Lovelace:2008tw} construct initial data and
perform short evolutions of BH-BH binaries with spins of 0.97.  More
recently, Lovelace et al~\cite{Lovelace:2010ne,Lovelace:2011nu}
compute complete inspiral/merger/ringdown simulations for equal mass
BH-BH binaries with aligned spins of 0.97 and anti-aligned spins of
0.95.  These simulations require delicate control of the excision
boundaries and demonstrate that the SpEC code has significantly
matured.  

The $\chi\!=\!0.97$ aligned spin simulation~\cite{Lovelace:2011nu} also carries
another new record: It is presently the longest published complete
BH-BH simulation, lasting about 25 orbits before merger and ringdown,
for an evolution time of 7000M.  The longest incomplete simulation 
(i.e. inspiral only) so
far is presented by Le Tiec et al.~\cite{LeTiec:2011bk}, lasting 34
orbits and 11,000M.  All these extremely long simulations were
obtained with SpEC.  The simulations presented in Ref.~\cite{LeTiec:2011bk}
range up to mass-ratio 8, all in excess of 20 orbits, and are used to
compare periastron advance in BH--BH binaries with several analytical 
approximation schemes.

The last year has also witnessed another increase in the largest known
black hole kick from inspiralling BH--BH\footnote{Interactions of
  high-velocity black holes result in even larger kicks~\cite{Healy:2008js}.}.  To remind the readers, in 2007 the ``BH
super-kick configuration'' was
discovered~\cite{Campanelli:2007ew,Gonzalez:2007hi,Campanelli:2007cga},
with BH kicks close to 4000km/sec.  In 2011, Lousto and
Zlochower~\cite{Lousto:2011kp} demonstrated that tilting the BH spins
from the super-kick configuration toward partial alignment with the
orbital angular momentum can increase the BH kicks to almost 5000km/sec.
Partial alignment of the BH spins with the orbital angular momentum
allows the BH's to spiral closer to each other, where the higher
velocities enhance the BH kick.

% There has also been progress on toward resolving the question
% whether toroidal event horizons can form in BH-BH mergers.  This
% question originated in theoretical and numerical work suggesting the
% existence of toroidal event
% horizons~\cite{Husa:1999nm,Diener:2003jc} during the dynamic,
% nonlinear merger of two black holes.  Recent work by Ponce et
% al~\cite{Poce:2010fq} and Cohen et
% al~\cite{Cohen:2011cf,Cohen:2008wa}

%%%%%%%%%%%%%%%%%%%%%%%%%%%%%%%%%%%%%%%%%%%%%%%%%%%%%%%%%%%%%%%%
\subsubsection{Improved techniques}
%\subsubsection{Controlling orbital eccentricity}
\label{sec:techniques}

{\bf Eccentricity removal} has been extended to precessing binaries:
Isolated BH-BH binaries originating from binary stars are expected to
have vanishing orbital eccentricity when they enter LIGO's frequency
band~\cite{Peters:1964zz,Peters:1963ux}.  To achieve low
eccentricity in a numerical simulation requires careful choices for
orbital frequency $\Omega_0$ and radial velocity $v_r$ of the
individual black holes (or, equivalently, tangential and radial linear momentum of the BH's).
%
% This motivates the need to
% numerically model low-eccentricity binaries.  BH-BH initial data
% contain parameters that determine the orbit of the binary, namely
% initial coordinate separation $d$, initial orbital frequency
% $\Omega_0$ (or tangential BH-momentum) and initial radial velocity
% $v_r$ (or radial BH momentum).  One can view $d$ as determining
% time-to-merger (number of orbits).  Having fixed $d$, initial orbital
% frequency $\Omega_0$ and radial velocity $v_r$ then determine the
% orbital eccentricity and phase of periastron.  Only one choice of
% $\Omega_0$ and $v_r$ will lead to a zero eccentricity inspiral.  
%
%There are two approaches exist to choose initial orbital parameters:
These parameters can be chosen based on post-Newtonian information, either in closed
form or by integrating post-Newtonian ordinary differential equations
for two point-masses starting at large separation.  The post-Newtonian
coordinates and velocities are then used in the construction of BH-BH 
initial--data. This approach~\cite{Husa:2007rh,Walther:2009ng} is
computationally inexpensive and achieves eccentricities of a
few $0.001$ for low-spin and comparable mass BH-BH, and somewhat
higher eccentricity for high spins and unequal mass BH-BH.

One can also adjust orbital parameters
iteratively~\cite{Boyle:2007ft,Pfeiffer:2007yz}: One begins with a
reasonable first guess for $\Omega_0$ and $v_r$ (perhaps based on
post-Newtonian information), evolves for about two orbits, analyzes
the orbital trajectories, and then adjusts $\Omega_0$ and $v_r$.  With
the adjusted initial orbital parameters, one constructs a new BH-BH
initial data set, and performs a new evolution.  This technique is
computationally more expensive but achieves eccentricities as
small as $\sim 10^{-5}$.  Recently, Buonnano et
al~\cite{Buonanno:2010yk} extended this technique to {\em precessing}
BH-BH systems.  Ref.~\cite{Buonanno:2010yk} performs a post-Newtonian
analysis demonstrating that spin impacts iterative eccentricity
removal only for eccentricities $10^{-4}\ldots 10^{-3}$ (the exact
bound depends on the BH-spins and the initial black hole separation).
Furthermore, Ref.~\cite{Buonanno:2010yk} demonstrates that $e\sim
10^{-4}$ can indeed be reached for several BH-BH configurations with
dimensionless spins of $0.5$.  This work uses the GH+Spectral approach
with the SpEC code.  Tichy et al~\cite{Tichy:2010qa} proposed an
alternative iterative technique within the BSSNOK+FD framework and
reach eccentricities of a few $\times 10^{-3}$.  Very recently
P{\"u}rrer et al.~\cite{Purrer:2012wy} study iterative eccentricity
removal for moving puncture simulations based on the
gravitational waveforms.

%%%%%%%%%%%%%%%%%%%%%%%%%%%%%%%%%%%%%%%%%%%%%%%%%%%%%%%%%%%%%%%%

{\bf Cauchy Characteristic
  extraction (CCE)}~\cite{Babiuc:2008qy,Reisswig:2009us,Babiuc:2010ze,Babiuc:2011qi}
extracts certain data from a standard 3+1 evolution code (as described
in Sec.~\ref{sec:NumericalMethods}) and evolves these data with a
separate characteristic code to future null infinity.  At
future null infinity, gravitational radiation is unambiguously
defined and so this technique promises gauge invariant waveforms,
improving on the more widely used technique to
extrapolate finite-radius waveforms to infinite extraction
radius~\cite{Boyle:2009vi}.  In recent papers, Babiuc et
al~\cite{Babiuc:2011qi,Babiuc:2010ze} improve the PITT CCE code,
and perform careful convergence tests.  This code is publicly
available as part of the Einstein
Toolkit~\cite{EinsteinToolkit,Loffler:2011ay}.  With the recent
improvements and public availability of the CCE code, I expect rapid
increase in the use of this post-processing tool.  
While CCE is gauge-invariant, it
requires initialization at the begin of the numerical simulation.
Bishop et al~\cite{Bishop:2011iu} investigate the impact of different
initialization strategies and found some impact onto the CCE waveform.  Furthermore, CCE cannot remove errors that
were introduced in the underlying 3+1 numerical evolution.  For
instance, if the outer boundary conditions of the 3+1 evolution admit
unphysical incoming radiation (or reflected outgoing radiation), then
such radiation will appear in the CCE waveform.

%%%%%%%%%%%%%%%%%%%%%%%%%%%%%%%%%%%%%%%%%%%%%%%%%%%%%%%%%%%%%%%%

\subsubsection{Code validation and consistency checks}

Several papers confirmed that BH-BH simulations indeed work as
expected thus further validating the numerical techniques:

Owen~\cite{Owen:2010vw,Owen:2009sb} (extending work by
Campanelli et al~\cite{Campanelli:2008dv}) analyzes carefully a SpEC
simulation of an equal-mass, non-spinning BH-BH.  He confirmed with a
multipolar analysis that the remnant black hole settles down to
Kerr with the correct quasi-normal mode falloffs.  The combination of
gauge-invariant quantitites with the high accuracy of the SpEC code
allows Owen to confirm agreement to many significant digits, usually to
better than 1 part in $10^5$ and sometimes to better than 1 part in $10^9$.
Refs.~\cite{Campanelli:2008dv,Owen:2010vw} also confirm that the
space-time of a black hole merger approaches Petrov type D at late
times after the BH-BH merger.

Hinder et al~\cite{Hinder:2011xx} perform a careful analysis of the
asymptotic fall-off of the Newman Penrose scalars,
revisiting~\cite{Pollney:2009ut}.  The fall-off rates agree with expectations,
\begin{equation}
\Psi_n \sim \frac{1}{r^{5-n}},
\end{equation}
where $n=0, \ldots 4$ labels the individual Newman-Penrose scalars,
$\Psi_0, \ldots, \Psi_4$.  

The NINJA-2 project~\cite{NinjaWebPage} collected about $40$ numerical
waveforms, with some configuration contributed multiple times by
different codes.  These duplicate waveforms allow consistency checks
between waveforms computed independently by different codes.  Overlap
calculations between (2,2) modes of hybridized waveforms show
disagreements broadly in line with the expected errors of the
different numerical codes and hybridization procedures.  Comparisons
of higher order modes have not yet been performed and are planned for
future work.  For details, see the separate NINJA-2 contribution to
the Amaldi proceedings~\cite{Ajith:2012tt}.

\subsubsection{Future improvements}

Several recent directions of research may have an impact on accuracy
and efficiency of BH-BH simulations in the future: The conformal
Z4-system~\cite{Alic:2011gg} combines features from BSSNOK with the
constraint damping of the generalized harmonic equations, resulting in
improved suppression of constraint violations (compared to BSSNOK).
Witek et al~\cite{Witek:2010es} developed a generalized BSSNOK system,
which encompases BSSNOK as a special case.  They find that their
extension improves numerical behavior of BSSN.  Existing BSSNOK codes
can easily be adopted to either the conformal Z4 system or the
generalized BSSNOK system.  Bona et al~\cite{Bona:2010is} presented a
Lagrangian for the Z4 system, which facilitates the use of symplectic
integrators for Einstein's equations.

Improved efficiency or reduced wall-clock run-time are in dire need,
given that a high-quality BH-BH simulation requires months to
complete.  At least two approaches are under development: Several
groups~\cite{Bruegmann:2011zj,MroueGpuTalk2010,EinsteinToolkit} work
on porting numerical relativity codes to graphical-processing-units
(GPUs), usually within the CUDA framework, opening up the possibility
that in the future GPUs may accelerate BH-BH simulations by a
significant factor.  Lau et al~\cite{Lau:2011we,Lau:2008fb} explore
novel time-stepping algorithms which circumvent the Courant timestep
limit, and promise the ability to take significantly larger timesteps
than current codes (see also~\cite{Hennig:2008af}).

%\newpage
%%%%%%%%%%%%%%%%%%%%%%%%%%%%%%%%%%%%%%%%%%%%%%%%%%%%%%%%%%%%%%%%
\subsection{BH-BH waveforms for gravitational wave astronomy}
\label{sec:BH-BH-GW}

The primary motivation for the intense activity in BH-BH research lies
in gravitational wave astronomy.  Gravitational wave detectors require
reasonably accurate waveform templates to detect GWs, and more
accurate waveforms to extract detailed knowledge about the source
properties (parameter estimation).  As discussed in detail in the
review~\cite{Ohme:2011rm}, one first performs BH-BH simulations at
discrete points in parameter space.  One then constructs analytical
waveform models (continuous in parameters) from these BH-BH
simulations, which are used for GW data-analysis purposes.  This
sequence of steps has been carried out several
times~\cite{Pan:2007nw,Buonanno:2007pf,Damour:2008te,Ajith:2009bn,Pan:2009wj,Santamaria:2010yb,Sturani:2010ju,Pan:2011gk},
based on different numerical simulations, different types of
analytical waveform models, and for different regions of parameter
space (again, we refer to Ref.~\cite{Ohme:2011rm} for details).

The NINJA projects~\cite{Aylott:2009ya,Ajith:2012tt,NinjaWebPage}
give a good sense of the rate of progress of BH--BH
simulations for GW-modeling. The initial NINJA
project~\cite{Aylott:2009ya} in 2008 consisted of about 20 numerical relativity
simulations lasting on average about 12 GW cycles before merger,
without any length- and accuracy-requirements.  The current NINJA-2
project~\cite{NinjaWebPage,Ajith:2012tt} collected about 40 waveforms
with on average about 20 GW cycles.  NINJA-2 also expands the coverage
of parameter space by including more spinning waveforms, 
the simulations are more accurate, and one carefully attaches post-Newtonian
inspirals to the numerical waveforms.  The second major ongoing effort, the
NR-AR collaboration~\cite{NRARWebPage} is assembling a yet larger
number of waveforms of average length of about 30 GW cycles, and with
yet higher accuracy than NINJA-2.  Each one  of these efforts
takes about two years to complete, demonstrating the high complexity
of computing, validating, and collecting high-quality BH-BH waveforms.

One obstacle toward computation of high-quality waveforms in a
shortage of experienced researchers capable of running BH-BH codes and
capable of improving the efficiency, robustness and automation of the
codes.  Furthermore, there were unexpected difficulties in scaling
BSSN-OK simulations to greater length and higher accuracy and for
the SpEC code to obtain robust and automatic mergers.  Finally,
computational resources are also constrained: As a rule of thumb, for
 ``easy'' parameter choices (moderate spins $\sim 0.5$,
moderate mass-ratios $\sim 2$, moderate length, $\sim 10$ orbits), a
single state-of-the-art BH-BH simulation requires on the order of
100,000 CPU-hours\footnote{The SpEC code is more efficient than
  the BSSN-OK codes.  But SpEC-simulations focus on longer and more
  accurate simulations, resulting in a CPU cost of the same order of magnitude.}.  This CPU-cost is of course
dependent on the length $T/M$ of the simulation (measured in units of
the total mass $M$).  For given symmetric mass-ratio $\nu$, initial
orbital frequency $\Omega_i$ or number $N$ of orbits to merger, lowest
order post-Newtonian expressions~\cite{Blanchet:2006zz} yield:
\begin{equation}\label{eq:TvsOmega}
\frac{T}{M}\approx \frac{5}{256}\, \nu^{-1} (M\Omega_i)^{-8/3},
\end{equation}
\begin{equation}\label{eq:TvsN}
\frac{T}{M}\approx 5\nu^{3/5}(2\pi N)^{8/5}.
\end{equation}
Halving the initial orbital frequency $\Omega_i$ or doubling the number of orbits
$N$ increases $T$ by a factor $\sim\! 6$ or $\sim\! 3$,
respectively.  The increase in CPU-cost is even higher, to
preserve phase-accuracy over the longer inspiral.  Higher mass-ratio and
higher spins increase the CPU-cost further.

Equations~(\ref{eq:TvsOmega}) and (\ref{eq:TvsN}) basically force a
trade-off between the length of each simulation and the number of 
simulations that can be performed with limited CPU resources.  
Reasonable accuracy for event-detection can be achieved
with $\sim10$ orbits~\cite{Ohme:2011zm,Hannam:2010ky}, however,
optimal parameter extraction requires numerical simulations starting at much
lower initial
frequency~\cite{Damour:2010zb,Boyle:2011dy,MacDonald:2011ne}.  The
low starting frequency is necessary because the 
accuracy of the hybrid waveform is primarily limited by the errors
of the 3.5-th order post-Newtonian waveforms that are attached {\em before} the start of the numerical
simulation.

\subsubsection{Exploration of parameter space}

The entire parameter space for BH-BH binaries is nine-dimensional:
mass-ratio, two spin-vectors, and two parameters related to
eccentricity (eccentricity and phase at periastron).  Essentially all
efforts to explore this parameter space so far have focused on
non-eccentric binaries, and even for non-eccentric binaries, only
the following low-dimensional subspaces have been covered in detail (see the NINJA-2
contribution~\cite{Ajith:2012tt} for details and references):
\begin{itemize}
\item Non-spinning binaries with mass-ratio $1\leq q\leq 10$.
\item Equal mass binaries with equal spins parallel to
 the orbital angular momentum.
\item Circular, non-precessing binaries form a three dimensional
  parameter space (mass-ratio and spin-magnitudes of the two spins
  parallel with the orbital angular momentum).  This space has not
  been covered extensively yet, with most efforts having been focused
  on the two one-dimensional subspaces just mentioned.
\end{itemize}

The vast seven-dimensional space of precessing binaries on circular
orbits has received so far surprisingly little attention: Campanelli
et al~\cite{Campanelli:2008nk} compute one waveform with mass-ratio
$q=1.25$, spins of 0.6 and 0.4, lasting about nine orbits.  A few
configurations are used to demonstrate BH--BH mergers in Szilagyi et
al.~\cite{Szilagyi:2009qz}, but without discussion of gravitational
waveforms.  Ref.~\cite{Buonanno:2010yk} perform several inspiral
simulations of precessing binaries to develop and test eccentricity
removal.  Sturani et al~\cite{Sturani:2010ju,Sturani:2010yv} perform
few-cycle long simulations, exploring a 1-dim line in the 7-dim
parameter space.  While these simulations have been used to construct
a phenomenological waveform-model of precessing
BH-BH~\cite{Sturani:2010ju,Sturani:2010yv}, the short length of numerical
simulations will severely limit the accuracy of this model.  Lousto
and Zlochower~\cite{Lousto:2011kp} perform 42 simulations to explore
black hole kicks with partial spin/orbit alignment.  Again, the short
length of the simulations (about five orbits) makes them unsuitable
for GW data-analysis purposes.  Recently, several papers 
discuss techniques to represent waveforms of precessing binaries by
expanding the waveforms in spherical harmonics with respect to a
time-dependent frame which is aligned with the instantaneous
radiation~\cite{Schmidt:2010it,O'Shaughnessy:2011fx,Boyle:2011gg} (of
which~\cite{Schmidt:2010it} presents a precessing $q\!=\!3$ BH--BH
simulation).  

As this short survey illustrates, several groups put their toe into the
ocean of precessing binaries, but nobody has seriously braved it yet.
The most significant current effort to explore precessing systems is the
 NR-AR collaboration~\cite{NRARWebPage}, which aims to construct
waveform models for precessing BH--BH systems.  While an intermediate
goal is to revisit the aligned spin case, this
collaboration will compute several high-quality precessing waveforms.

%%%%%%%%%%%%%%%%%%%%%%%%%%%%%%%%%%%%%%%%%%%%%%%%%%%%%%%%%%%%%%%%
\subsection{BH--BH in non-vacuum}

Fully relativistic hydro-dynamics simulations of BH-BH in gaseous
environment have continued.   Bode et al~\cite{Bode:2009mt} and
Bogdanovic et al~\cite{Bogdanovic:2010he} consider radiatively
inefficient accretion flow on merging BH-BH binaries, modeled as a
BH-BH embedded in a cloud of gas with Gaussian density profile
centered on the center of mass.  Farris et al~\cite{Farris:2009mt}
investigate the BH-BH analog of Bondi-accretion, embedding the BH-BH
in ambient gas of constant density\footnote{Zanotti et
al~\cite{Zanotti:2011mb} studied the standard Bondi accretion onto 
a single black hole with general relativistic radiation-hydrodynamics.}.

Accretion disks are also under continued investigation.
Refs.~\cite{Anderson:2009fa,Megevand:2009yx,Zanotti:2010xs} 
study the behavior of a circumbinary disk {\em after} the BH-BH
merger, when the disk responds to the remnant black hole with reduced mass 
(due to energy loss through gravitational waves) and with non-zero velocity relative to the center of mass of the
accretion disk (due to black hole kicks).
More recently, Bode et al.~\cite{Bode:2011tq} and Farris et al.~\cite{Farris:2011vx} investigate BH-BH
binaries with a circumbinary disk. 

%EM counterparts of recoiling BH~\cite{Megevand:2009yzZanotti:2010xs}\\
%Perturbed disks get shocked~\cite{Megevand:2009yx}\\
%Post-merger EM emission from disks~\cite{Anderson:2009fa}

Two groups are investigating the effects of electro-magnetic fields
surrounding BH-BH binaries.  This work started with solving the vaccum
Maxwell equations coupled to GR by Palenzuela et
al~\cite{Palenzuela:2009hx,Palenzuela:2009yr} and Moesta et
al.~\cite{Mosta:2009rr}.  More recently  a force-free treatment of the electro-magnetic fields was presented in
Refs.~\cite{Palenzuela:2010nf,Palenzuela:2010xn,Neilsen:2010ax,%
  Palenzuela:2011es,Moesta:2011bn}.  The force-free approximation
assumes the presence of a tenuous plasma which shortens out any
electric field parallel to the magnetic field lines.  Such a plasma is
produced by pair-production and forms the basis for the
Blandford-Znajek process~\cite{Blandford:1977ds} to extract energy
from rotating black holes.  Palenzuela et al~\cite{Palenzuela:2010nf},
in particular, demonstrate how a merger of a spinning BH-BH can result
in a total of three jets: During the inspiral, one jet associated
with each BH; and after the merger, a third jet associated with
the remnant BH.  Moesta et al~\cite{Moesta:2011bn} consider the strength of
this beamed emission relative to the uniform emission that also
accompanies mergers of BH-BH in these cases.

%On the detectability~\cite{Moesta:2011bn}\\
%Vaccuum EM counterpartds of BBH~\cite{Mosta:2009rr}\\
%\\
%Robustness of BZ~\cite{Palenzuela:2011es}\\
%Boosting jet-power~\cite{Neilsen:2010ax}\\
%Magnetospheres of BH systems in FF~\cite{Palenzuela:2010xn}\\
%Dual jets from BBH~\cite{Palenzuela:2010nf}\\
%possible EM counterparts~\cite{Palenzuela:2009hx}\\
%Stirring, not shaking~\cite{Palenzuela:2009yr}

%%%%%%%%%%%%%%%%%%%%%%%%%%%%%%%%%%%%%%%%%%%%%%%%%%%%%%%%%%%%%%%%

\section{BH--NS \& NS--NS systems}
\label{sec:BH-NS/NS-NS}

We now turn to simulations with at least one Neutron star
(NS).  These simulations have a more varied set of goals than the
BH-BH simulations, among them:
\begin{enumerate}
\item Compute gravitational waveforms to aid GW detectors.  This
  requires simulations covering many inspiral orbits at high
  phase-accuracy, a significant challenge for hydro-codes which
  are typically less accurate than vacuum BH-BH codes.
\item Investigate properties of the merger: In BH-NS binaries, does the Neutron star get disrupted?  Does an accretion disk form, and how large is it?  Is the binary a viable progenitor for a short gamma ray burst?
\item Are these mergers a viable source for r-process
  elements~\cite{1999ApJ...525L.121F}?  This requires the formation of
  unbound ejecta which return material into the
  interstellar medium.
\end{enumerate}

A variety of physical effects may affect evolution of these binaries
and all of these effect must be studied numerically:
impact of mass-ratio, effect of equation of state, effect of
magnetic fields, effect of nuclear physics, effect of neutrino
transport.  For BH-NS systems, one must also investigate the effect of
the BH spin (both in magnitude and direction).  Neutron stars are
generally assumed to be very slowly spinning, albeit some recent
work approaches simulations of spinning
NS~\cite{Baumgarte:2009fw,Tichy:2011gw}.  Several different groups
investigate BH--NS and NS--NS systems, with each group generally
focusing on a a subset of these physical effects, as detailed in
Secs.~\ref{sec:BH--NS} and~\ref{sec:NS--NS}.

The numerical techniques for BH--NS and NS--NS combine a solver for the Einstein
equations with appropriate solvers for the matter fields.  Most commonly used is the BSSNOK evolution
system for gravity, combined with high-resolution shock capturing
techniques for relativistic hydrodynamics, discretized with
adaptive-mesh refinement finite-differences.  The SpEC collaboration
combines a spectral treatment of the gravity sector (similar to their BH-BH
simulations) with a treatment of matter on a separate Cartesian
finite-difference grid that covers only the region of space in which
matter is present and that moves along with the Neutron stars.
Some recent papers describing numerical techniques and individual codes are Refs.~\cite{Yamamoto:2008js,%SACRA
Giacomazzo:2007ti,%WhiskeyMHD
Baiotti:2010ka,%WhiskeyvsSacra
Duez:2008rb,%
East:2011aa,%East 
Thierfelder:2011yi,%
Etienne:2011re,Etienne:2010ui,%
Anderson:2006ay%
}.
Initial data for compact object binaries with Neutron stars are described in Refs.~\cite{Uryu:2011ky,Kyutoku:2009sp,Uryu:2009ye,Grandclement:2006ht,Taniguchi:2007xm,Foucart:2008qt}.

Detailed investigations into accuracy of BH--NS and NS--NS simulations
are given in
Refs.~\cite{Baiotti:2009gk,Foucart:2011mz,Bernuzzi:2011aq}.  Broadly
speaking, preserving low-phase errors during inspiral is difficult
and simulations with Neutron stars are less accurate than their BH--BH
counterparts.  No current code is clearly superior; in particular, the
advantage of {\tt SpEC} for vacuum simulations does not carry over and
{\tt SpEC}--hydro-simulations are of comparable accuracy than other
codes.  Resolving disk-dynamics is another numerically challenging aspect,
especially when the disks are long-lived.  Finally, MHD simulations
require a lower cutoff on the matter density.
Because gamma-ray-bursts involve outflows in regions of low Baryon
density, the MHD caveats might become important (the IMEX
treatment of Palenzuela et al~\cite{Palenzuela:2008sf} is a recent
alternative).

%%%%%%%%%%%%%%%%%%%%%%%%%%%%%%%%%%%%%%%%%%%%%%%%%%%%%%%%%%%%%%%%
\subsection{BH--NS}
\label{sec:BH--NS}

One key objective of the study of BH--NS binaries 
lies in finding the region of parameter space where
the Neutron star is tidally disrupted.  In this
case, an accretion disk forms and the system is considered 
a possible
candidate for a gamma-ray burst (although disruption may not be necessary for a gamma-ray burst, see~\cite{Hansen:2000am,McWilliams:2011zi}).  Moreover, the gravitational wave
signature changes dramatically at disruption:  Once the Neutron star is
disrupted and spread into an accretion disk, GW emission is
drastically reduced.

Tidal disruption depends primarily on three parameters.  (1) The
compaction of the Neutron star: more compact stars are harder to
disrupt.  (2) The mass of the black hole:  more massive black
holes have weaker tidal forces, reducing the tendency of
disruption.  (3) The distance between Neutron star and black hole:
For spinning black holes, the innermost stable orbit of corotating
geodesics moves inward; therefore, for BH--NS systems with the BH spin
aligned with the orbital angular momentum, the orbit of the Neutron star
will reach a smaller distance, thus increasing the tendency of
disruption. 

A significant amount of attention
has been devoted on determining the parameter space in which accretion
disks form and on the properties of the formed accretion disks.  Let us
summarize some recent work: Shibata et al.~\cite{Shibata:2009cn}
investigate polytropic Neutron stars orbiting non-spinning black
holes.  For mass-ratios $q\!=\!M_{\rm BH}/N_{\rm NS}\!=\!1.5$ and $3$,
disruption occurs, but not so for $q\!=\!5$.
Kyutoku et al.~\cite{Kyutoku:2011vz} consider BH's
with spin aligned or antialigned with the orbital angular momentum
with spin magnitudes up to $\chi_{\rm BH}\equiv S_{\rm BH}/M_{\rm
  BH}^2=0.5$.  For low mass-ratio $q\leq 3$ the Neutron star
disrupts.  For $q\!=\!5$, disruption occurs only for prograde BH-spin
(this work uses a piecewise polytropic equation of state, which is
fitted to physical equation of states).  The remnant disks have mass of $\sim 0.1M_\odot$ and this work (as the other
references mentioned) confirm that indeed NS disruption leads to a
cutoff in the gravitational wave spectrum.  A related study by Chawla
et al.~\cite{Chawla:2010sw} agrees with the results of
\cite{Kyutoku:2011vz}:  Focusing on mass-ratio $q\!=\!5$,
aligned co-rotating BH with spin  of $\chi_{\rm BH}\!=\!S_{\rm BH}/M_{\rm BH}^2\!=\!0.5$, and a
polytropic equation of state, Chawla et al find that an accretion disk
forms, with essentially all material gravitationally bound.  
Ref.~\cite{Chawla:2010sw} also incorporates magnetic fields, and for
field-strengths up to $10^{12}G$, the effect of the magnetic field was
found to be marginal.
  
BH--NS binaries with relatively low mass-ratio $q\lesssim 5$ are numerically easier
to handle and are interesting because these systems form accretion
disks easily.  However, population synthesis suggests that black holes
are generally more massive~\cite{Belczynski:2010tb}.  Foucart et
al.~\cite{Foucart:2011mz} perform simulations with more
massive black holes, $M_{\rm BH}\!=\!10M_{\odot}$ and mass-ratio $q\!=\!7$.
For a non-spinning black hole, or moderately corotating black hole, no
accretion disk forms at $q\!=\!7$.  Only at high spins $\chi_{\rm BH}\gtrsim 0.7$
does an accretion disk develop.  These simulations investigate BHs with
spins as large as $\chi_{\rm BH}\!=\!0.9$, the largest to date.

To close our summary on the effect of BH spin and mass-ratio, we note
that all simulations mentioned so far have the BH spin parallel to the
orbital angular momentum.  This assumption is removed by 
Foucart et al.~\cite{Foucart:2010eq} who vary the spin direction of
the BH.  Specifically, the angle $\theta$ between
BH spin and orbital angular momentum is varied between $0^\circ$ and $80^\circ$
(for $q\!=\!3$, BH spin $\chi_{\rm BH}\!=\!0.5$ and a polytropic equation of
state).  Angles $\theta>40^\circ$ result in a reduction in disk-mass by about a
factor of 2.  For the same mass-ratio,
Ref.~\cite{Foucart:2010eq} also investigate aligned BH spins with
magnitude $\chi_{\rm BH}=0.9$.  Consistent with expectations, such a
high co-rotating BH spin coupled with the low mass-ratio $q=3$ results in 
a very large disk with mass approaching  $\sim 0.4 M_{\rm NS}$.

A second focal point for recent work was the impact of the equation of
state (EOS) of the Neutron star matter.  Duez et al.~\cite{Duez:2009yy}
investigate polytropic equations of state with two different
polytropic indices ($\Gamma\!=\!2$ and $\Gamma\!=\!2.75$), and the Shen EOS
with two treatments of the electron fraction.  The binary
has mass-ratio $q\!=\!3$ and an aligned BH spin $\chi_{\rm BH}\!=\!0.5$.  Duez et al find that more compact Neutron stars emit stronger gravitational
waves and result in smaller disk-masses.  In particular, tidal tails
depend on the EOS.  Kyotuko~et al.~\cite{Kyutoku:2010zd} investigate
piece-wise polytropic equations of
state~\cite{Read:2008iy,Ozel:2009da}.  They find that less compact
Neutron stars tend to disrupt at a larger separation, and that
disk-mass (and the characteristic GW cutoff frequency) correlate with
compaction of the Neutron star.

Recently,  the first MHD simulations of BH--NS binaries were performed.  Chawla
et al.~\cite{Chawla:2010sw} consider field-strengths of $10^{12}G$ and
find no appreciable difference to non-magnetic binaries, whereas Etienne et
al.~\cite{Etienne:2011ea} report that a magnetic field of $10^{17}G$
results in a increased disk-mass (no evidence for collimated outflows
were observed, although ~\cite{Etienne:2011ea} points out that higher
resolution simulations would be needed, that follow the post-merger
dynamics for a longer period of time).

Stephens et al.~\cite{Stephens:2011as} and East et
al.~\cite{East:2011xa} consider hyperbolic encounters of a 
Neutron star with a black hole (with relative 
velocity $1000$km/sec at large distance).  
Refs.~\cite{Stephens:2011as,East:2011xa}
vary the equation of state, impact parameter, and BH spin.  By far the
biggest effect on the results has the impact parameter: Depending on
the periastron distance, the NS can be disrupted in the first
approach, periodic mass-transfer can occur, or the NS can pass the BH
unharmed.

%%%%%%%%%%%%%%%%%%%%%%%%%%%%%%%%%%%%%%%%%%%%%%%%%%%%%%%%%%%%%%%%

\subsection{NS--NS}
\label{sec:NS--NS}

Binary Neutron stars have seen an equal amount of activity lately as
BH--NS binaries.  The themes are quite similar to BH--NS, with 
intense efforts of the research groups to extend the range of
included physical effects.

Equation of state effects are exhaustively explored by
Hotokezaka et al.~\cite{Hotokezaka:2011dh}.  They simulate six
different EOS's (parametrized by piecewise
polytropic~\cite{Read:2008iy,Ozel:2009da}) for three different
NS-masses each and classify the results into prompt collapse to a
BH, shortlived hypermassive Neutron star, and long-lived
hypermassive Neutron stars.  Hotokezaka et al report torus masses of
up to $0.1M_\odot$ around the newly formed black hole.  Sekiguchi et
al~\cite{Sekiguchi:2011mc} explore an equation of state with a
phase-transition to hyperons.  This causes substantial differences in
dynamics, observable in gravitational waves.

When a NS--NS merger results in a hypermassive Neutron star (as
opposed to prompt collapse), shocks during the merger will heat the
remnant NS to high temperatures, and neutrino cooling will become
important.  Therefore, neutrino
cooling must be modeled in order to determine reliably the timescale
on which the hypermassive NS cools and collapses to a BH when the thermal pressure becomes
insufficient to support the star.  Sekiguchi et
al.~\cite{Sekiguchi:2011zd} perform the first simulation of this
process, incorporating neutrino cooling with a leakage scheme, and
using a finite-temperature Shen EOS.  They report the neutrino luminosities
for the simulated NS--NS mergers, and find indeed that the lifetime of
low-mass hypermassive NS depends on the neutrino cooling.

MHD simulations have also been improved.  Giacomazzo et
al~\cite{Giacomazzo:2010bx} and Rezzolla et al~\cite{Rezzolla:2011da}
focus on very long simulations of the post-merger accretion disk.
While Ref.~\cite{Giacomazzo:2010bx} focuses on disk dynamics,
Ref.~\cite{Rezzolla:2011da} reports evidence of electromagnetic
collimation along the rotation axis of the accretion disk.  This is
the first claim of the ''missing link'' that connects the post-merger
accretion disk with the eventual launching of the jets that power
gamma ray bursts.  Two other research groups have published
simulations of merging magnetized NS--NS
systems~\cite{Liu:2008xy,Anderson:2008zp}, and both these groups have
since published improved techniques to handle magnetic
fields~\cite{Etienne:2011re,Etienne:2010ui,Liebling:2010bn,Palenzuela:2008sf}.
It will be very interesting to see whether the impressive and
important results of Ref.~\cite{Rezzolla:2011da} can be confirmed by
these groups.

Eccentric NS--NS systems were also studied for the first time.
Gold et al.~\cite{Gold:2011df} find that eccentric mergers result in
larger disks, and that periastron passage leads to excitation of
f-modes in the Neutron stars.

Finally, recently work has begun to compare NS--NS inspiral
simulations with post-Newtonian approximations, and to fit analytical
waveform models to the numerical NS--NS inspiral simulations.  Such
work is important for gravitational wave data-analysis, because
complete, phase-accurate waveforms (ranging from early post-Newtonian
inspiral into merger) promise the most accurate data-analysis for
gravitational waves from NS--NS.  Several NS--NS simulations were performed for these purposes all about 10 orbits long.  Bernuzzi et
al.~\cite{Bernuzzi:2011aq} performs a comparison with TaylorT4 and
finds significant phase-differences (about 1 GW cylce), which cannot
be explained by known tidal corrections.  Baiotti et
al~\cite{Baiotti:2010xh,Baiotti:2011am} investigate TaylorT4 and EOB
models.  Without free parameters, EOB and TaylorT4 both deviate by
about one GW cycle from the numerical NS--NS simulation; in contrast, if the
EOB model is amended with one free fitting parameter (parameterizing
the strength of higher order tidal effects), the NS--NS simulation can
be fitted with an error of only 0.24 radians. 

Read et al.~\cite{Read:2009yp} and Lackey et al.\cite{Lackey:2011vz}
analyze large sets of NS--NS simulations to determine what
information about the equation of state can be extracted from future
gravitational wave observations.  The best-constrained parameter will be
the compactness of the Neutron star.

\section{Conclusion}
\label{sec:discussion}

The recent progress in simulations of BH--BH, BH--NS and NS--NS
systems is spectacular.  What are the challenges going forward?

For BH--BH, the numerical methods are in good shape.  Very challenging
simulations were successfully performed as described in
Sec.~\ref{sec:recent-advances}, pushing large mass-ratio, large spins, and
the
number of GW-cycles.  The open question is whether the numerical
techniques can handle a combination of these properties, e.g. high
spin ($\gtrsim 0.9$) and high mass-ratio ($\gtrsim 4$), and how well
accuracy holds up when the number of GW cycles in the simulation is
increased.  Besides these unexplored regions of parameter space, the
main challenge forward is the systematic exploration of the vast
parameter space of possible binaries, at sufficient accuracy and
length to allow gravitational wave detectors to reach optimal
sensitivity and optimal accuracy in parameter estimation.  The
computational cost of BH--BH simulations, the possibility that the
numerical waveforms may have to be much longer than current
simulations for {\em optimal} parameter estimation, and the sheer size
of the parameter space will necessitate compromises, in length of the
simulations or in parameter space coverage (or both).  The severity of
these compromises will only be known after an initial exploration of
the precessing parameter space and after first attempts to fit
analytical waveform models.  The ease of fitting (currently unknown)
will determine how many simulations are needed.

Efficiency improvements in the numerical codes will furthermore
determine how quickly the parameter space can be sampled and how
quickly on-demand simulations can be performed (e.g. in response to
gravitational wave detectors observing a tentative event).  Such
efficiency improvements may come from novel computer algorithms like
implicit time-stepping~\cite{Lau:2011we,Lau:2008fb} or from utilizing
novel computing paradigms like graphical processing units.

For BH--NS and NS--NS systems, the current frontier is exploration of
all relevant physical effects.  Given the complexity of the
simulations and the varied micro-physics, it is imperative that
different groups perform similar simulations to cross-check.  Once
qualitative features are explored, the field will turn towards
quantitative sampling of the parameter space, with systematic and
careful calculation of gravitational waveforms.

Numerical and implementation issues of compact object binaries seem a
problem of the past, as attested by the stunning progress in numerical
relativity.  The deeper insight into formulation of the equations, and
experience of what works and what does not makes it now possible to
consider entirely different numerical algorithms, like discontinuous
Galerkin
methods~\cite{hesthaven08:_nodal_discon_galer_method,Radice:2011qr},
moving Voronoi meshes~\cite{Duffell:2011bc,Springel:2011yv}, or novel
time-stepping techniques~\cite{Lau:2011we}.

\section*{Acknowledgments}

I thank the organizers of the 9th Eduardo Amaldi conference and the
5th Numerical Relativity---Data Analysis meeting for organizing two
extremely stimulating conferences.   I thank Matt Duez and Francois Foucart 
for illuminating
discussions, and gratefully acknowledge support from NSERC of Canada,
the Canada Chairs Program and the Canadian Institute for Advanced
Research.

 \section*{References}
\bibliographystyle{iopart-num}
\bibliography{refs}

\providecommand{\newblock}{}
\begin{thebibliography}{100}
\expandafter\ifx\csname url\endcsname\relax
  \def\url#1{{\tt #1}}\fi
\expandafter\ifx\csname urlprefix\endcsname\relax\def\urlprefix{URL }\fi
\providecommand{\eprint}[2][]{\url{#2}}

\bibitem{Lousto:2011kp}
Lousto C~O and Zlochower Y 2011 {\em Phys.Rev.Lett.\/} {\bf 107} 231102
  (\textit{Preprint} \eprint{1108.2009})

\bibitem{Cohen:2011cf}
Cohen M~I, Kaplan J~D and Scheel M~A 2012 {\em Phys.Rev.\/} {\bf D85} 024031
  (\textit{Preprint} \eprint{1110.1668})

\bibitem{Cohen:2008wa}
Cohen M~I, Pfeiffer H~P and Scheel M~A 2009 {\em Class.Quant.Grav.\/} {\bf 26}
  035005 (\textit{Preprint} \eprint{0809.2628})

\bibitem{Ponce:2010fq}
Ponce M, Lousto C and Zlochower Y 2011 {\em Class.Quant.Grav.\/} {\bf 28}
  145027 (\textit{Preprint} \eprint{1008.2761})

\bibitem{Duez:2009yz}
Duez M~D 2010 {\em Class.Quant.Grav.\/} {\bf 27} 114002 (\textit{Preprint}
  \eprint{0912.3529})

\bibitem{lrr-2011-6}
Shibata M and Taniguchi K 2011 {\em Living Reviews in Relativity\/} {\bf 14}
  \urlprefix\url{http://www.livingreviews.org/lrr-2011-6}

\bibitem{Abbott:2007kv}
Abbott B {\em et~al.\/} (LIGO Scientific) 2009 {\em Rept. Prog. Phys.\/} {\bf
  72} 076901 (\textit{Preprint} \eprint{0711.3041})

\bibitem{Shoemaker:aLIGO}
Shoemaker D (the Advanced LIGO Team) 2009 {Advanced LIGO Reference Design}
  {[LIGO-M060056]}

\bibitem{2010CQGra..27h4006H}
{Harry} G~M and {the LIGO Scientific Collaboration} 2010 {\em Class. Quant.
  Grav.\/} {\bf 27} 084006

\bibitem{Accadia:2011zz}
Accadia T, Acernese F, Antonucci F, Astone P, Ballardin G {\em et~al.\/} 2011
  {\em Class. Quant. Grav.\/} {\bf 28} 114002

\bibitem{aVIRGO}
{The Virgo Collaboration} 2009 {Advanced Virgo Baseline Design}
  {[VIR-0027A-09]}

\bibitem{Somiya:2011me}
Somiya K (for the LCGT Collaboration) 2011  (\textit{Preprint}
  \eprint{1111.7185})

\bibitem{Centrella:2010mx}
Centrella J, Baker J~G, Kelly B~J and van Meter J~R 2010 {\em Rev.Mod.Phys.\/}
  {\bf 82} 3069 (\textit{Preprint} \eprint{1010.5260})

\bibitem{McWilliams:2010iq}
McWilliams S~T 2011 {\em Class.Quant.Grav.\/} {\bf 28} 134001
  (\textit{Preprint} \eprint{1012.2872})

\bibitem{Baumgarte-Shapiro-Book:2010}
Baumgarte T~W and Shapiro S~L 2010 {\em Numerical Relativity: Solving
  Einstein's Equations on the Computer\/} (Cambridge University Press)

\bibitem{Witek:2010xi}
Witek H, Zilhao M, Gualtieri L, Cardoso V, Herdeiro C {\em et~al.\/} 2010 {\em
  Phys.Rev.\/} {\bf D82} 104014 (\textit{Preprint} \eprint{1006.3081})

\bibitem{Lehner:2010pn}
Lehner L and Pretorius F 2010 {\em Phys.Rev.Lett.\/} {\bf 105} 101102
  (\textit{Preprint} \eprint{1006.5960})

\bibitem{Healy:2011ef}
Healy J, Bode T, Haas R, Pazos E, Laguna P {\em et~al.\/} 2011
  (\textit{Preprint} \eprint{1112.3928})

\bibitem{Paschalidis:2011ww}
Paschalidis V, Halataei S~M, Shapiro S~L and Sawicki I 2011 {\em
  Class.Quant.Grav.\/} {\bf 28} 085006 (\textit{Preprint} \eprint{1103.0984})

\bibitem{Pretorius:2005gq}
Pretorius F 2005 {\em Phys.Rev.Lett.\/} {\bf 95} 121101 (\textit{Preprint}
  \eprint{gr-qc/0507014})

\bibitem{Campanelli:2005dd}
Campanelli M, Lousto C, Marronetti P and Zlochower Y 2006 {\em
  Phys.Rev.Lett.\/} {\bf 96} 111101 (\textit{Preprint} \eprint{gr-qc/0511048})

\bibitem{Baker:2005vv}
Baker J~G, Centrella J, Choi D~I, Koppitz M and van Meter J 2006 {\em
  Phys.Rev.Lett.\/} {\bf 96} 111102 (\textit{Preprint} \eprint{gr-qc/0511103})

\bibitem{Nakamura:1987zz}
Nakamura T, Oohara K and Kojima Y 1987 {\em Prog. Theor. Phys. Suppl.\/} {\bf
  90} 1--218

\bibitem{Shibata:1995we}
Shibata M and Nakamura T 1995 {\em Phys. Rev\/} {\bf D52} 5428--5444

\bibitem{Baumgarte:1998te}
Baumgarte T~W and Shapiro S~L 1999 {\em Phys. Rev.\/} {\bf D59} 024007
  (\textit{Preprint} \eprint{gr-qc/9810065})

\bibitem{Friedrich1985}
Friedrich H 1985 {\em Commun. Math. Phys.\/} {\bf 100} 525--543

\bibitem{Pretorius:2004jg}
Pretorius F 2005 {\em Class.Quant.Grav.\/} {\bf 22} 425--452 (\textit{Preprint}
  \eprint{gr-qc/0407110})

\bibitem{Lindblom:2005qh}
Lindblom L, Scheel M~A, Kidder L~E, Owen R and Rinne O 2006 {\em
  Class.Quant.Grav.\/} {\bf 23} S447--S462 (\textit{Preprint}
  \eprint{gr-qc/0512093})

\bibitem{Sperhake:2008ga}
Sperhake U, Cardoso V, Pretorius F, Berti E and Gonzalez J~A 2008 {\em Phys.
  Rev. Lett.\/} {\bf 101} 161101 (\textit{Preprint} \eprint{0806.1738})

\bibitem{Pretorius:2007jn}
Pretorius F and Khurana D 2007 {\em Class.Quant.Grav.\/} {\bf 24} S83--S108
  (\textit{Preprint} \eprint{gr-qc/0702084})

\bibitem{Ramazanoglu:2010aj}
Ramazanoglu F~M and Pretorius F 2010 {\em Class.Quant.Grav.\/} {\bf 27} 245027
  (\textit{Preprint} \eprint{1009.1440})

\bibitem{O'Murchadha:1974nc}
O'Murchadha N and York J~W 1974 {\em Phys.Rev.\/} {\bf D10} 428--436

\bibitem{Pfeiffer:2002iy}
Pfeiffer H~P and York James~W J 2003 {\em Phys.Rev.\/} {\bf D67} 044022
  (\textit{Preprint} \eprint{gr-qc/0207095})

\bibitem{Bowen:1980yu}
Bowen J~M and York James~W J 1980 {\em Phys.Rev.\/} {\bf D21} 2047--2056

\bibitem{York:1998hy}
York James~W J 1999 {\em Phys.Rev.Lett.\/} {\bf 82} 1350--1353
  (\textit{Preprint} \eprint{gr-qc/9810051})

\bibitem{Brandt:1997tf}
Brandt S and Bruegmann B 1997 {\em Phys.Rev.Lett.\/} {\bf 78} 3606--3609
  (\textit{Preprint} \eprint{gr-qc/9703066})

\bibitem{Caudill:2006hw}
Caudill M, Cook G~B, Grigsby J~D and Pfeiffer H~P 2006 {\em Phys.Rev.\/} {\bf
  D74} 064011 (\textit{Preprint} \eprint{gr-qc/0605053})

\bibitem{Cook:2004kt}
Cook G~B and Pfeiffer H~P 2004 {\em Phys.Rev.\/} {\bf D70} 104016
  (\textit{Preprint} \eprint{gr-qc/0407078})

\bibitem{Cook:2001wi}
Cook G~B 2002 {\em Phys.Rev.\/} {\bf D65} 084003 (\textit{Preprint}
  \eprint{gr-qc/0108076})

\bibitem{Ansorg:2004ds}
Ansorg M, Bruegmann B and Tichy W 2004 {\em Phys.Rev.\/} {\bf D70} 064011
  (\textit{Preprint} \eprint{gr-qc/0404056})

\bibitem{Pfeiffer:2002wt}
Pfeiffer H~P, Kidder L~E, Scheel M~A and Teukolsky S~A 2003 {\em
  Comput.Phys.Commun.\/} {\bf 152} 253--273 (\textit{Preprint}
  \eprint{gr-qc/0202096})

\bibitem{Husa:2007rh}
Husa S, Hannam M, Gonzalez J~A, Sperhake U and Bruegmann B 2008 {\em
  Phys.Rev.\/} {\bf D77} 044037 (\textit{Preprint} \eprint{0706.0904})

\bibitem{Pfeiffer:2007yz}
Pfeiffer H~P, Brown D~A, Kidder L~E, Lindblom L, Lovelace G {\em et~al.\/} 2007
  {\em Class.Quant.Grav.\/} {\bf 24} S59--S82 (\textit{Preprint}
  \eprint{gr-qc/0702106})

\bibitem{Buonanno:2010yk}
Buonanno A, Kidder L~E, Mroue A~H, Pfeiffer H~P and Taracchini A 2011 {\em
  Phys.Rev.\/} {\bf D83} 104034 (\textit{Preprint} \eprint{1012.1549})

\bibitem{Bona:1997hp}
Bona C, Masso J, Seidel E and Stela J 1997 {\em Phys. Rev.\/} {\bf D56}
  3405--3415 (\textit{Preprint} \eprint{gr-qc/9709016})

\bibitem{Alcubierre:2002kk}
Alcubierre M {\em et~al.\/} 2003 {\em Phys. Rev.\/} {\bf D67} 084023
  (\textit{Preprint} \eprint{gr-qc/0206072})

\bibitem{vanMeter:2006vi}
van Meter J~R, Baker J~G, Koppitz M and Choi D~I 2006 {\em Phys.Rev.\/} {\bf
  D73} 124011 (\textit{Preprint} \eprint{gr-qc/0605030})

\bibitem{Szilagyi:2009qz}
Szilagyi B, Lindblom L and Scheel M~A 2009 {\em Phys.Rev.\/} {\bf D80} 124010
  (\textit{Preprint} \eprint{0909.3557})

\bibitem{Scheel:2006gg}
Scheel M~A, Pfeiffer H~P, Lindblom L, Kidder L~E, Rinne O {\em et~al.\/} 2006
  {\em Phys.Rev.\/} {\bf D74} 104006 (\textit{Preprint} \eprint{gr-qc/0607056})

\bibitem{Rinne:2007ui}
Rinne O, Lindblom L and Scheel M~A 2007 {\em Class.Quant.Grav.\/} {\bf 24}
  4053--4078 (\textit{Preprint} \eprint{0704.0782})

\bibitem{Husa:2007hp}
Husa S, Gonzalez J~A, Hannam M, Bruegmann B and Sperhake U 2008 {\em
  Class.Quant.Grav.\/} {\bf 25} 105006 (\textit{Preprint} \eprint{0706.0740})

\bibitem{Zlochower:2005bj}
Zlochower Y, Baker J, Campanelli M and Lousto C 2005 {\em Phys.Rev.\/} {\bf
  D72} 024021 (\textit{Preprint} \eprint{gr-qc/0505055})

\bibitem{Brown:2007pg}
Brown J, Sarbach O, Schnetter E, Tiglio M, Diener P {\em et~al.\/} 2007 {\em
  Phys.Rev.\/} {\bf D76} 081503 (\textit{Preprint} \eprint{0707.3101})

\bibitem{Brown:2008sb}
Brown J, Diener P, Sarbach O, Schnetter E and Tiglio M 2009 {\em Phys.Rev.\/}
  {\bf D79} 044023 (\textit{Preprint} \eprint{0809.3533})

\bibitem{Etienne:2007hr}
Etienne Z~B, Faber J~A, Liu Y~T, Shapiro S~L and Baumgarte T~W 2007 {\em
  Phys.Rev.\/} {\bf D76} 101503 (\textit{Preprint} \eprint{0707.2083})

\bibitem{Lovelace:2011nu}
Lovelace G, Boyle M, Scheel M~A and Szilagyi B 2012 {\em Class.Quant.Grav.\/}
  {\bf 29} 045003 (\textit{Preprint} \eprint{1110.2229})

\bibitem{Hannam:2009ib}
Hannam M, Husa S and Murchadha N~O 2009 {\em Phys.Rev.\/} {\bf D80} 124007
  (\textit{Preprint} \eprint{0908.1063})

\bibitem{Goodale02a}
Goodale T, Allen G, Lanfermann G, Mass{\'o} J, Radke T, Seidel E and Shalf J
  2003 {\em Vector and Parallel Processing -- VECPAR'2002, 5th International
  Conference, Lecture Notes in Computer Science\/} (Berlin: Springer) pp
  197--227

\bibitem{cactus}
Cactus computational toolkit \urlprefix\url{http://www.cactuscode.org/}

\bibitem{Loffler:2011ay}
Loffler F, Faber J, Bentivegna E, Bode T, Diener P {\em et~al.\/} 2011
  (\textit{Preprint} \eprint{1111.3344})

\bibitem{Schnetter:2003rb}
Schnetter E, Hawley S~H and Hawke I 2004 {\em Class.Quant.Grav.\/} {\bf 21}
  1465--1488 (\textit{Preprint} \eprint{gr-qc/0310042})

\bibitem{carpet}
Schnetter E Carpet: A mesh refinement driver for cactus
  \urlprefix\url{http://www.carpetcode.org/}

\bibitem{Thornburg:2003sf}
Thornburg J 2004 {\em Class.Quant.Grav.\/} {\bf 21} 743--766 (\textit{Preprint}
  \eprint{gr-qc/0306056})

\bibitem{Ajith:2012tt}
Ajith P, Boyle M, Brown D~A, Brugmann B, Buchman L~T {\em et~al.\/} 2012
  (\textit{Preprint} \eprint{1201.5319})

\bibitem{SpECwebsite}
\url{http://www.black-holes.org/SpEC.html}

\bibitem{Owen:2010fa}
Owen R, Brink J, Chen Y, Kaplan J~D, Lovelace G {\em et~al.\/} 2011 {\em
  Phys.Rev.Lett.\/} {\bf 106} 151101 (\textit{Preprint} \eprint{1012.4869})

\bibitem{Nichols:2011pu}
Nichols D~A, Owen R, Zhang F, Zimmerman A, Brink J {\em et~al.\/} 2011 {\em
  Phys.Rev.\/} {\bf D84} 124014 (\textit{Preprint} \eprint{1108.5486})

\bibitem{Muller:2010zze}
Muller D, Grigsby J and Bruegmann B 2010 {\em Phys.Rev.\/} {\bf D82} 064004
  (\textit{Preprint} \eprint{1003.4681})

\bibitem{Schnetter:2010cz}
Schnetter E 2010 {\em Class.Quant.Grav.\/} {\bf 27} 167001 (\textit{Preprint}
  \eprint{1003.0859})

\bibitem{Alic:2010wu}
Alic D, Rezzolla L, Hinder I and Mosta P 2010 {\em Class.Quant.Grav.\/} {\bf
  27} 245023 (\textit{Preprint} \eprint{1008.2212})

\bibitem{Chu:2009md}
Chu T, Pfeiffer H~P and Scheel M~A 2009 {\em Phys.Rev.\/} {\bf D80} 124051
  (\textit{Preprint} \eprint{0909.1313})

\bibitem{Sperhake:2011ik}
Sperhake U, Cardoso V, Ott C~D, Schnetter E and Witek H 2011 {\em Phys.Rev.\/}
  {\bf D84} 084038 (\textit{Preprint} \eprint{1105.5391})

\bibitem{Lousto:2010ut}
Lousto C~O and Zlochower Y 2011 {\em Phys.Rev.Lett.\/} {\bf 106} 041101
  (\textit{Preprint} \eprint{1009.0292})

\bibitem{Nakano:2011pb}
Nakano H, Zlochower Y, Lousto C~O and Campanelli M 2011 {\em Phys. Rev.\/} {\bf
  D84} 124006 (\textit{Preprint} \eprint{1108.4421})

\bibitem{Lousto:2010qx}
Lousto C~O, Nakano H, Zlochower Y and Campanelli M 2010 {\em Phys. Rev.\/} {\bf
  D82} 104057 (\textit{Preprint} \eprint{1008.4360})

\bibitem{Lousto:2010tb}
Lousto C~O, Nakano H, Zlochower Y and Campanelli M 2010 {\em Phys. Rev.
  Lett.\/} {\bf 104} 211101 (\textit{Preprint} \eprint{1001.2316})

\bibitem{Dain:2008ck}
Dain S, Lousto C~O and Zlochower Y 2008 {\em Phys.Rev.\/} {\bf D78} 024039
  (\textit{Preprint} \eprint{0803.0351})

\bibitem{Lovelace:2008tw}
Lovelace G, Owen R, Pfeiffer H~P and Chu T 2008 {\em Phys.Rev.\/} {\bf D78}
  084017 (\textit{Preprint} \eprint{0805.4192})

\bibitem{Lovelace:2010ne}
Lovelace G, Scheel M and Szilagyi B 2011 {\em Phys.Rev.\/} {\bf D83} 024010
  (\textit{Preprint} \eprint{1010.2777})

\bibitem{LeTiec:2011bk}
Le~Tiec A, Mroue A~H, Barack L, Buonanno A, Pfeiffer H~P {\em et~al.\/} 2011
  {\em Phys.Rev.Lett.\/} {\bf 107} 141101 (\textit{Preprint}
  \eprint{1106.3278})

\bibitem{Healy:2008js}
Healy J, Herrmann F, Hinder I, Shoemaker D~M, Laguna P {\em et~al.\/} 2009 {\em
  Phys.Rev.Lett.\/} {\bf 102} 041101 (\textit{Preprint} \eprint{0807.3292})

\bibitem{Campanelli:2007ew}
Campanelli M, Lousto C~O, Zlochower Y and Merritt D 2007 {\em Astrophys.J.\/}
  {\bf 659} L5--L8 (\textit{Preprint} \eprint{gr-qc/0701164})

\bibitem{Gonzalez:2007hi}
Gonzalez J, Hannam M, Sperhake U, Bruegmann B and Husa S 2007 {\em
  Phys.Rev.Lett.\/} {\bf 98} 231101 (\textit{Preprint} \eprint{gr-qc/0702052})

\bibitem{Campanelli:2007cga}
Campanelli M, Lousto C~O, Zlochower Y and Merritt D 2007 {\em Phys.Rev.Lett.\/}
  {\bf 98} 231102 (\textit{Preprint} \eprint{gr-qc/0702133})

\bibitem{Peters:1964zz}
Peters P 1964 {\em Phys.Rev.\/} {\bf 136} B1224--B1232

\bibitem{Peters:1963ux}
Peters P and Mathews J 1963 {\em Phys.Rev.\/} {\bf 131} 435--439

\bibitem{Walther:2009ng}
Walther B, Bruegmann B and Mueller D 2009 {\em Phys.Rev.\/} {\bf D79} 124040
  (\textit{Preprint} \eprint{0901.0993})

\bibitem{Boyle:2007ft}
Boyle M, Brown D~A, Kidder L~E, Mroue A~H, Pfeiffer H~P {\em et~al.\/} 2007
  {\em Phys.Rev.\/} {\bf D76} 124038 (\textit{Preprint} \eprint{0710.0158})

\bibitem{Tichy:2010qa}
Tichy W and Marronetti P 2011 {\em Phys.Rev.\/} {\bf D83} 024012
  (\textit{Preprint} \eprint{1010.2936})

\bibitem{Purrer:2012wy}
Purrer M, Husa S and Hannam M 2012  (\textit{Preprint} \eprint{1203.4258})

\bibitem{Babiuc:2008qy}
Babiuc M, Bishop N, Szilagyi B and Winicour J 2009 {\em Phys.Rev.\/} {\bf D79}
  084011 (\textit{Preprint} \eprint{0808.0861})

\bibitem{Reisswig:2009us}
Reisswig C, Bishop N, Pollney D and Szilagyi B 2009 {\em Phys.Rev.Lett.\/} {\bf
  103} 221101 (\textit{Preprint} \eprint{0907.2637})

\bibitem{Babiuc:2010ze}
Babiuc M, Szilagyi B, Winicour J and Zlochower Y 2011 {\em Phys.Rev.\/} {\bf
  D84} 044057 (\textit{Preprint} \eprint{1011.4223})

\bibitem{Babiuc:2011qi}
Babiuc M, Winicour J and Zlochower Y 2011 {\em Class.Quant.Grav.\/} {\bf 28}
  134006 (\textit{Preprint} \eprint{1106.4841})

\bibitem{Boyle:2009vi}
Boyle M and Mroue A~H 2009 {\em Phys.Rev.\/} {\bf D80} 124045
  (\textit{Preprint} \eprint{0905.3177})

\bibitem{EinsteinToolkit}
\url{http://einsteintoolkit.org}

\bibitem{Bishop:2011iu}
Bishop N, Pollney D and Reisswig C 2011 {\em Class.Quant.Grav.\/} {\bf 28}
  155019 (\textit{Preprint} \eprint{1101.5492})

\bibitem{Owen:2010vw}
Owen R 2010 {\em Phys.Rev.\/} {\bf D81} 124042 (\textit{Preprint}
  \eprint{1004.3768})

\bibitem{Owen:2009sb}
Owen R 2009 {\em Phys.Rev.\/} {\bf D80} 084012 (\textit{Preprint}
  \eprint{0907.0280})

\bibitem{Campanelli:2008dv}
Campanelli M, Lousto C~O and Zlochower Y 2009 {\em Phys.Rev.\/} {\bf D79}
  084012 (\textit{Preprint} \eprint{0811.3006})

\bibitem{Hinder:2011xx}
Hinder I, Wardell B and Bentivegna E 2011 {\em Phys.Rev.\/} {\bf D84} 024036
  (\textit{Preprint} \eprint{1105.0781})

\bibitem{Pollney:2009ut}
Pollney D, Reisswig C, Dorband N, Schnetter E and Diener P 2009 {\em
  Phys.Rev.\/} {\bf D80} 121502 (\textit{Preprint} \eprint{0910.3656})

\bibitem{NinjaWebPage}
\url{https://www.ninja-project.org}

\bibitem{Alic:2011gg}
Alic D, Bona-Casas C, Bona C, Rezzolla L and Palenzuela C 2011
  (\textit{Preprint} \eprint{1106.2254})

\bibitem{Witek:2010es}
Witek H, Hilditch D and Sperhake U 2011 {\em Phys.Rev.\/} {\bf D83} 104041
  (\textit{Preprint} \eprint{1011.4407})

\bibitem{Bona:2010is}
Bona C, Bona-Casas C and Palenzuela C 2010 {\em Phys.Rev.\/} {\bf D82} 124010
  (\textit{Preprint} \eprint{1008.0747})

\bibitem{Bruegmann:2011zj}
Bruegmann B 2011  (\textit{Preprint} \eprint{1104.3408})

\bibitem{MroueGpuTalk2010}
Mrou\'e A~H 2010 Binary black hole simulations using {CUDA} {NVIDIA} {GPU}
  Technology Conference

\bibitem{Lau:2011we}
Lau S~R, Lovelace G and Pfeiffer H~P 2011 {\em Phys.Rev.\/} {\bf D84} 084023
  (\textit{Preprint} \eprint{1105.3922})

\bibitem{Lau:2008fb}
Lau S~R, Pfeiffer H~P and Hesthaven J~S 2008 {\em Comput.Phys.Commun.\/}
  (\textit{Preprint} \eprint{0808.2597})

\bibitem{Hennig:2008af}
Hennig J and Ansorg M 2009 {\em J.Hyperbol.Diff.Equat.\/} {\bf 6} 161
  (\textit{Preprint} \eprint{0801.1455})

\bibitem{Ohme:2011rm}
Ohme F 2011  (\textit{Preprint} \eprint{1111.3737})

\bibitem{Pan:2007nw}
Pan Y, Buonanno A, Baker J~G, Centrella J, Kelly B~J {\em et~al.\/} 2008 {\em
  Phys.Rev.\/} {\bf D77} 024014 (\textit{Preprint} \eprint{0704.1964})

\bibitem{Buonanno:2007pf}
Buonanno A, Pan Y, Baker J~G, Centrella J, Kelly B~J {\em et~al.\/} 2007 {\em
  Phys.Rev.\/} {\bf D76} 104049 (\textit{Preprint} \eprint{0706.3732})

\bibitem{Damour:2008te}
Damour T, Nagar A, Hannam M, Husa S and Bruegmann B 2008 {\em Phys.Rev.\/} {\bf
  D78} 044039 (\textit{Preprint} \eprint{0803.3162})

\bibitem{Ajith:2009bn}
Ajith P, Hannam M, Husa S, Chen Y, Bruegmann B {\em et~al.\/} 2011 {\em
  Phys.Rev.Lett.\/} {\bf 106} 241101 (\textit{Preprint} \eprint{0909.2867})

\bibitem{Pan:2009wj}
Pan Y, Buonanno A, Buchman L~T, Chu T, Kidder L~E {\em et~al.\/} 2010 {\em
  Phys.Rev.\/} {\bf D81} 084041 (\textit{Preprint} \eprint{0912.3466})

\bibitem{Santamaria:2010yb}
Santamaria L, Ohme F, Ajith P, Bruegmann B, Dorband N {\em et~al.\/} 2010 {\em
  Phys.Rev.\/} {\bf D82} 064016 (\textit{Preprint} \eprint{1005.3306})

\bibitem{Sturani:2010ju}
Sturani R, Fischetti S, Cadonati L, Guidi G, Healy J {\em et~al.\/} 2010
  (\textit{Preprint} \eprint{1012.5172})

\bibitem{Pan:2011gk}
Pan Y, Buonanno A, Boyle M, Buchman L~T, Kidder L~E {\em et~al.\/} 2011 {\em
  Phys.Rev.\/} {\bf D84} 124052 26 pages, 25 figures, published Phys. Rev. D
  version (\textit{Preprint} \eprint{1106.1021})

\bibitem{Aylott:2009ya}
Aylott B, Baker J~G, Boggs W~D, Boyle M, Brady P~R {\em et~al.\/} 2009 {\em
  Class.Quant.Grav.\/} {\bf 26} 165008 (\textit{Preprint} \eprint{0901.4399})

\bibitem{NRARWebPage}
\url{https://www.ninja-project.org/doku.php?id=nrar:home}

\bibitem{Blanchet:2006zz}
Blanchet L 2006 {\em Living Rev.Rel.\/} {\bf 9} 4

\bibitem{Ohme:2011zm}
Ohme F, Hannam M and Husa S 2011 {\em Phys.Rev.\/} {\bf D84} 064029
  (\textit{Preprint} \eprint{1107.0996})

\bibitem{Hannam:2010ky}
Hannam M, Husa S, Ohme F and Ajith P 2010 {\em Phys.Rev.\/} {\bf D82} 124052
  (\textit{Preprint} \eprint{1008.2961})

\bibitem{Damour:2010zb}
Damour T, Nagar A and Trias M 2011 {\em Phys.Rev.\/} {\bf D83} 024006
  (\textit{Preprint} \eprint{1009.5998})

\bibitem{Boyle:2011dy}
Boyle M 2011 {\em Phys.Rev.\/} {\bf D84} 064013 (\textit{Preprint}
  \eprint{1103.5088})

\bibitem{MacDonald:2011ne}
MacDonald I, Nissanke S, Pfeiffer H~P and Pfeiffer H~P 2011 {\em
  Class.Quant.Grav.\/} {\bf 28} 134002 (\textit{Preprint} \eprint{1102.5128})

\bibitem{Campanelli:2008nk}
Campanelli M, Lousto C~O, Nakano H and Zlochower Y 2009 {\em Phys. Rev.\/} {\bf
  D79} 084010 (\textit{Preprint} \eprint{0808.0713})

\bibitem{Sturani:2010yv}
Sturani R, Fischetti S, Cadonati L, Guidi G, Healy J {\em et~al.\/} 2010 {\em
  J.Phys.Conf.Ser.\/} {\bf 243} 012007 (\textit{Preprint} \eprint{1005.0551})

\bibitem{Schmidt:2010it}
Schmidt P, Hannam M, Husa S and Ajith P 2011 {\em Phys.Rev.\/} {\bf D84} 024046
  (\textit{Preprint} \eprint{1012.2879})

\bibitem{O'Shaughnessy:2011fx}
O'Shaughnessy R, Vaishnav B, Healy J, Meeks Z and Shoemaker D 2011 {\em
  Phys.Rev.\/} {\bf D84} 124002 (\textit{Preprint} \eprint{1109.5224})

\bibitem{Boyle:2011gg}
Boyle M, Owen R and Pfeiffer H~P 2011 {\em Phys.Rev.\/} {\bf D84} 124011
  (\textit{Preprint} \eprint{1110.2965})

\bibitem{Bode:2009mt}
Bode T, Haas R, Bogdanovic T, Laguna P and Shoemaker D 2010 {\em
  Astrophys.J.\/} {\bf 715} 1117--1131 (\textit{Preprint} \eprint{0912.0087})

\bibitem{Bogdanovic:2010he}
Bogdanovic T, Bode T, Haas R, Laguna P and Shoemaker D 2011 {\em
  Class.Quant.Grav.\/} {\bf 28} 094020 (\textit{Preprint} \eprint{1010.2496})

\bibitem{Farris:2009mt}
Farris B~D, Liu Y~T and Shapiro S~L 2010 {\em Phys.Rev.\/} {\bf D81} 084008
  (\textit{Preprint} \eprint{0912.2096})

\bibitem{Zanotti:2011mb}
Zanotti O, Roedig C, Rezzolla L and Del~Zanna L 2011 {\em
  Mon.Not.Roy.Astron.Soc.\/} {\bf 417} 2899--2915 (\textit{Preprint}
  \eprint{1105.5615})

\bibitem{Anderson:2009fa}
Anderson M, Lehner L, Megevand M and Neilsen D 2010 {\em Phys.Rev.\/} {\bf D81}
  044004 (\textit{Preprint} \eprint{0910.4969})

\bibitem{Megevand:2009yx}
Megevand M, Anderson M, Frank J, Hirschmann E~W, Lehner L {\em et~al.\/} 2009
  {\em Phys.Rev.\/} {\bf D80} 024012 (\textit{Preprint} \eprint{0905.3390})

\bibitem{Zanotti:2010xs}
Zanotti O, Rezzolla L, Del~Zanna L and Palenzuela C 2010 {\em
  Astron.Astrophys.\/} {\bf 523} A8 (\textit{Preprint} \eprint{1002.4185})

\bibitem{Bode:2011tq}
Bode T, Bogdanovic T, Haas R, Healy J, Laguna P {\em et~al.\/} 2012 {\em
  Astrophys.J.\/} {\bf 744} 45 (\textit{Preprint} \eprint{1101.4684})

\bibitem{Farris:2011vx}
Farris B~D, Liu Y~T and Shapiro S~L 2011 {\em Phys.Rev.\/} {\bf D84} 024024
  (\textit{Preprint} \eprint{1105.2821})

\bibitem{Palenzuela:2009hx}
Palenzuela C, Lehner L and Yoshida S 2010 {\em Phys.Rev.\/} {\bf D81} 084007
  (\textit{Preprint} \eprint{0911.3889})

\bibitem{Palenzuela:2009yr}
Palenzuela C, Anderson M, Lehner L, Liebling S~L and Neilsen D 2009 {\em
  Phys.Rev.Lett.\/} {\bf 103} 081101 (\textit{Preprint} \eprint{0905.1121})

\bibitem{Mosta:2009rr}
Mosta P, Palenzuela C, Rezzolla L, Lehner L, Yoshida S {\em et~al.\/} 2010 {\em
  Phys.Rev.\/} {\bf D81} 064017 (\textit{Preprint} \eprint{0912.2330})

\bibitem{Palenzuela:2010nf}
Palenzuela C, Lehner L and Liebling S~L 2010 {\em Science\/} {\bf 329} 927
  (\textit{Preprint} \eprint{1005.1067})

\bibitem{Palenzuela:2010xn}
Palenzuela C, Garrett T, Lehner L and Liebling S~L 2010 {\em Phys.Rev.\/} {\bf
  D82} 044045 (\textit{Preprint} \eprint{1007.1198})

\bibitem{Neilsen:2010ax}
Neilsen D, Lehner L, Palenzuela C, Hirschmann E~W, Liebling S~L {\em et~al.\/}
  2011 {\em Proc.Nat.Acad.Sci.\/} {\bf 108} 12641--12646 (\textit{Preprint}
  \eprint{1012.5661})

\bibitem{Palenzuela:2011es}
Palenzuela C, Bona C, Lehner L and Reula O 2011 {\em Class.Quant.Grav.\/} {\bf
  28} 134007 (\textit{Preprint} \eprint{1102.3663})

\bibitem{Moesta:2011bn}
Moesta P, Alic D, Rezzolla L, Zanotti O and Palenzuela C 2011
  (\textit{Preprint} \eprint{1109.1177})

\bibitem{Blandford:1977ds}
Blandford R and Znajek R 1977 {\em Mon.Not.Roy.Astron.Soc.\/} {\bf 179}
  433--456

\bibitem{1999ApJ...525L.121F}
{Freiburghaus} C, {Rosswog} S and {Thielemann} F~K 1999 {\em Astrophys. J.
  Lett.\/} {\bf 525} L121--L124

\bibitem{Baumgarte:2009fw}
Baumgarte T~W and Shapiro S~L 2009 {\em Phys.Rev.\/} {\bf D80} 064009
  (\textit{Preprint} \eprint{0909.0952})

\bibitem{Tichy:2011gw}
Tichy W 2011 {\em Phys.Rev.\/} {\bf D84} 024041 (\textit{Preprint}
  \eprint{1107.1440})

\bibitem{Yamamoto:2008js}
Yamamoto T, Shibata M and Taniguchi K 2008 {\em Phys.Rev.\/} {\bf D78} 064054
  (\textit{Preprint} \eprint{0806.4007})

\bibitem{Giacomazzo:2007ti}
Giacomazzo B and Rezzolla L 2007 {\em Class.Quant.Grav.\/} {\bf 24} S235--S258
  (\textit{Preprint} \eprint{gr-qc/0701109})

\bibitem{Baiotti:2010ka}
Baiotti L, Shibata M and Yamamoto T 2010 {\em Phys.Rev.\/} {\bf D82} 064015
  (\textit{Preprint} \eprint{1007.1754})

\bibitem{Duez:2008rb}
Duez M~D {\em et~al.\/} 2008 {\em Phys. Rev.\/} {\bf D78} 104015
  (\textit{Preprint} \eprint{0809.0002})

\bibitem{East:2011aa}
East W~E, Pretorius F and Stephens B~C 2011  (\textit{Preprint}
  \eprint{1112.3094})

\bibitem{Thierfelder:2011yi}
Thierfelder M, Bernuzzi S and Bruegmann B 2011 {\em Phys.Rev.\/} {\bf D84}
  044012 (\textit{Preprint} \eprint{1104.4751})

\bibitem{Etienne:2011re}
Etienne Z~B, Paschalidis V, Liu Y~T and Shapiro S~L 2012 {\em Phys.Rev.\/} {\bf
  D85} 024013 (\textit{Preprint} \eprint{1110.4633})

\bibitem{Etienne:2010ui}
Etienne Z~B, Liu Y~T and Shapiro S~L 2010 {\em Phys.Rev.\/} {\bf D82} 084031
  (\textit{Preprint} \eprint{1007.2848})

\bibitem{Anderson:2006ay}
Anderson M, Hirschmann E, Liebling S~L and Neilsen D 2006 {\em
  Class.Quant.Grav.\/} {\bf 23} 6503--6524 (\textit{Preprint}
  \eprint{gr-qc/0605102})

\bibitem{Uryu:2011ky}
Uryu K and Tsokaros A 2012 {\em Phys.Rev.\/} {\bf D85} 064014 revised version
  with a new title, 29 pages (\textit{Preprint} \eprint{1108.3065})

\bibitem{Kyutoku:2009sp}
Kyutoku K, Shibata M and Taniguchi K 2009 {\em Phys. Rev.\/} {\bf D79} 124018
  (\textit{Preprint} \eprint{0906.0889})

\bibitem{Uryu:2009ye}
Uryu K, Limousin F, Friedman J~L, Gourgoulhon E and Shibata M 2009 {\em
  Phys.Rev.\/} {\bf D80} 124004 (\textit{Preprint} \eprint{0908.0579})

\bibitem{Grandclement:2006ht}
Grandclement P 2006 {\em Phys.Rev.\/} {\bf D74} 124002 (\textit{Preprint}
  \eprint{gr-qc/0609044})

\bibitem{Taniguchi:2007xm}
Taniguchi K, Baumgarte T~W, Faber J~A and Shapiro S~L 2007 {\em Phys.Rev.\/}
  {\bf D75} 084005 (\textit{Preprint} \eprint{gr-qc/0701110})

\bibitem{Foucart:2008qt}
Foucart F, Kidder L~E, Pfeiffer H~P and Teukolsky S~A 2008 {\em Phys. Rev.\/}
  {\bf D77} 124051 (\textit{Preprint} \eprint{0804.3787})

\bibitem{Baiotti:2009gk}
Baiotti L, Giacomazzo B and Rezzolla L 2009 {\em Class.Quant.Grav.\/} {\bf 26}
  114005 (\textit{Preprint} \eprint{0901.4955})

\bibitem{Foucart:2011mz}
Foucart F, Duez M~D, Kidder L~E, Scheel M~A, Szilagyi B {\em et~al.\/} 2012
  {\em Phys.Rev.\/} {\bf D85} 044015 11 pages, 11 figures - Updated to match
  published version (\textit{Preprint} \eprint{1111.1677})

\bibitem{Bernuzzi:2011aq}
Bernuzzi S, Thierfelder M and Bruegmann B 2011  (\textit{Preprint}
  \eprint{1109.3611})

\bibitem{Palenzuela:2008sf}
Palenzuela C, Lehner L, Reula O and Rezzolla L 2009 {\em Mon. Not. Roy. Astron.
  Soc.\/} {\bf 394} 1727--1740 (\textit{Preprint} \eprint{0810.1838})

\bibitem{Hansen:2000am}
Hansen B~M and Lyutikov M 2001 {\em Mon.Not.Roy.Astron.Soc.\/} {\bf 322} 695
  (\textit{Preprint} \eprint{astro-ph/0003218})

\bibitem{McWilliams:2011zi}
McWilliams S~T and Levin J 2011 {\em Astrophys.J.\/} {\bf 742} 90
  (\textit{Preprint} \eprint{1101.1969})

\bibitem{Shibata:2009cn}
Shibata M, Kyutoku K, Yamamoto T and Taniguchi K 2009 {\em Phys.Rev.\/} {\bf
  D79} 044030 (\textit{Preprint} \eprint{0902.0416})

\bibitem{Kyutoku:2011vz}
Kyutoku K, Okawa H, Shibata M and Taniguchi K 2011 {\em Phys. Rev.\/} {\bf D84}
  064018 (\textit{Preprint} \eprint{1108.1189})

\bibitem{Chawla:2010sw}
Chawla S {\em et~al.\/} 2010 {\em Phys. Rev. Lett.\/} {\bf 105} 111101
  (\textit{Preprint} \eprint{1006.2839})

\bibitem{Belczynski:2010tb}
Belczynski K, Dominik M, Bulik T, O'Shaughnessy R, Fryer C {\em et~al.\/} 2010
  (\textit{Preprint} \eprint{1004.0386})

\bibitem{Foucart:2010eq}
Foucart F, Duez M~D, Kidder L~E and Teukolsky S~A 2011 {\em Phys. Rev.\/} {\bf
  D83} 024005 (\textit{Preprint} \eprint{1007.4203})

\bibitem{Duez:2009yy}
Duez M~D, Foucart F, Kidder L~E, Ott C~D and Teukolsky S~A 2010 {\em Class.
  Quant. Grav.\/} {\bf 27} 114106 (\textit{Preprint} \eprint{0912.3528})

\bibitem{Kyutoku:2010zd}
Kyutoku K, Shibata M and Taniguchi K 2010 {\em Phys. Rev.\/} {\bf D82} 044049
  (\textit{Preprint} \eprint{1008.1460})

\bibitem{Read:2008iy}
Read J~S, Lackey B~D, Owen B~J and Friedman J~L 2009 {\em Phys.Rev.\/} {\bf
  D79} 124032 (\textit{Preprint} \eprint{0812.2163})

\bibitem{Ozel:2009da}
Ozel F and Psaltis D 2009 {\em Phys.Rev.\/} {\bf D80} 103003 (\textit{Preprint}
  \eprint{0905.1959})

\bibitem{Etienne:2011ea}
Etienne Z~B, Liu Y~T, Paschalidis V and Shapiro S~L 2011  (\textit{Preprint}
  \eprint{1112.0568})

\bibitem{Stephens:2011as}
Stephens B~C, East W~E and Pretorius F 2011 {\em Astrophys.J.\/} {\bf 737} L5
  (\textit{Preprint} \eprint{1105.3175})

\bibitem{East:2011xa}
East W~E, Pretorius F and Stephens B~C 2011  (\textit{Preprint}
  \eprint{1111.3055})

\bibitem{Hotokezaka:2011dh}
Hotokezaka K, Kyutoku K, Okawa H, Shibata M and Kiuchi K 2011 {\em Phys.Rev.\/}
  {\bf D83} 124008 (\textit{Preprint} \eprint{1105.4370})

\bibitem{Sekiguchi:2011mc}
Sekiguchi Y, Kiuchi K, Kyutoku K and Shibata M 2011 {\em Phys.Rev.Lett.\/} {\bf
  107} 211101 (\textit{Preprint} \eprint{1110.4442})

\bibitem{Sekiguchi:2011zd}
Sekiguchi Y, Kiuchi K, Kyutoku K and Shibata M 2011 {\em Phys.Rev.Lett.\/} {\bf
  107} 051102 (\textit{Preprint} \eprint{1105.2125})

\bibitem{Giacomazzo:2010bx}
Giacomazzo B, Rezzolla L and Baiotti L 2011 {\em Phys.Rev.\/} {\bf D83} 044014
  (\textit{Preprint} \eprint{1009.2468})

\bibitem{Rezzolla:2011da}
Rezzolla L, Giacomazzo B, Baiotti L, Granot J, Kouveliotou C {\em et~al.\/}
  2011 {\em Astrophys.J.\/} {\bf 732} L6 (\textit{Preprint} \eprint{1101.4298})

\bibitem{Liu:2008xy}
Liu Y~T, Shapiro S~L, Etienne Z~B and Taniguchi K 2008 {\em Phys. Rev.\/} {\bf
  D78} 024012 (\textit{Preprint} \eprint{0803.4193})

\bibitem{Anderson:2008zp}
Anderson M {\em et~al.\/} 2008 {\em Phys. Rev. Lett.\/} {\bf 100} 191101
  (\textit{Preprint} \eprint{0801.4387})

\bibitem{Liebling:2010bn}
Liebling S~L, Lehner L, Neilsen D and Palenzuela C 2010 {\em Phys. Rev.\/} {\bf
  D81} 124023 (\textit{Preprint} \eprint{1001.0575})

\bibitem{Gold:2011df}
Gold R, Bernuzzi S, Thierfelder M, Bruegmann B and Pretorius F 2011
  (\textit{Preprint} \eprint{1109.5128})

\bibitem{Baiotti:2010xh}
Baiotti L, Damour T, Giacomazzo B, Nagar A and Rezzolla L 2010 {\em
  Phys.Rev.Lett.\/} {\bf 105} 261101 (\textit{Preprint} \eprint{1009.0521})

\bibitem{Baiotti:2011am}
Baiotti L, Damour T, Giacomazzo B, Nagar A and Rezzolla L 2011 {\em
  Phys.Rev.\/} {\bf D84} 024017 (\textit{Preprint} \eprint{1103.3874})

\bibitem{Read:2009yp}
Read J~S, Markakis C, Shibata M, Uryu K, Creighton J~D {\em et~al.\/} 2009 {\em
  Phys.Rev.\/} {\bf D79} 124033 (\textit{Preprint} \eprint{0901.3258})

\bibitem{Lackey:2011vz}
Lackey B~D, Kyutoku K, Shibata M, Brady P~R and Friedman J~L 2012 {\em
  Phys.Rev.\/} {\bf D85} 044061 (\textit{Preprint} \eprint{1109.3402})

\bibitem{hesthaven08:_nodal_discon_galer_method}
Hesthaven J~S and Warburton T 2008 {\em Nodal Discontinuous Galerkin Methods:
  Algorithms, Analysis, and Applications\/} (Springer)

\bibitem{Radice:2011qr}
Radice D and Rezzolla L 2011 {\em Phys.Rev.\/} {\bf D84} 024010
  (\textit{Preprint} \eprint{1103.2426})

\bibitem{Duffell:2011bc}
Duffell P~C and MacFadyen A~I 2011 {\em Astrophys.J.Suppl.\/} {\bf 197} 15
  (\textit{Preprint} \eprint{1104.3562})

\bibitem{Springel:2011yv}
Springel V 2011  (\textit{Preprint} \eprint{1109.2218})

\end{thebibliography}

\end{document}